\documentclass[conference]{IEEEtran}
\IEEEoverridecommandlockouts
\usepackage{cite}
\usepackage{amsmath,amssymb,amsfonts}
\usepackage{algorithmic}
\usepackage{graphicx}
\usepackage{textcomp}
\usepackage{xcolor}
\usepackage{xr}
\usepackage{hyperref}
\def\BibTeX{{\rm B\kern-.05em{\sc i\kern-.025em b}\kern-.08em
    T\kern-.1667em\lower.7ex\hbox{E}\kern-.125emX}}

\begin{document}

\title{Users volatility on Reddit and Voat}

\author{\IEEEauthorblockN{Niccol{\`o} Di Marco \IEEEauthorrefmark{1},
Matteo Cinelli \IEEEauthorrefmark{2},
Shayan Alipour \IEEEauthorrefmark{2}, \\
and
 Walter Quattrociocchi \IEEEauthorrefmark{2},~\IEEEmembership{Fellow,~IEEE}}
\IEEEauthorblockA{\IEEEauthorrefmark{1} Department of Mathematics and Computer Science, University of Florence, Florence, Italy}
\IEEEauthorblockA{\IEEEauthorrefmark{2} Center of Data Science and Complexity for Society,\\ Department of Computer Science, Sapienza University of Rome, Rome, Italy}
% \IEEEauthorblockA{\IEEEauthorrefmark{3} Enrico Fermi Research Center, Rome, Italy}
\thanks{Corresponding author: W. Quattrociocchi (email: walter.quattrociocchi@gmail.com).}}

\maketitle

\begin{abstract}

Social media platforms are like giant arenas where users can rely on different content and express their opinions through likes, comments, and shares. However, do users welcome different perspectives or only listen to their preferred narratives?
This paper examines how users explore the digital space and allocate their attention among communities on two social networks, Voat and Reddit. 
By analysing a massive dataset of about 215 million comments posted by about 16 million users on Voat and Reddit in 2019 we find that most users tend to explore new communities at a decreasing rate, meaning they have a limited set of preferred groups they visit regularly. Moreover, we provide evidence that preferred communities of users tend to cover similar topics throughout the year. We also find that communities have a high turnover of users, meaning that users come and go frequently showing a high volatility that strongly departs from a null model simulating users' behaviour. 
% Finally, we use information theory tools to measure users' loyalty to their communities, finding that users behave very differently from random behaviour.

% Our results suggest that users are curious but need to be more adventurous in their exploration of the digital space. They tend to form attachments to some communities and abandon others. This leads to high volatility and low loyalty in most communities. We discuss the implications of these findings for social media platforms and their users.

\end{abstract}


\section{Introduction}

 Social media exposes us to vast information, creating an excess of choices for content consumption. For instance, the volume of videos uploaded to platforms like YouTube or TikTok exceeds the time available for human viewing. To cope with this information overload, in which providers vie for our attention~\cite{simon1971designing}, we employ cognitive and cultural/attitudinal mechanisms that emerge in various contexts, such as social interactions~\cite{SBH,Dunbar1992,cinelli2021} communication~\cite{Saramki2014,Gonalves2011, Miritello2013} and mobility~\cite{Alessandretti2018}. These mechanisms, although helpful, can also have detrimental results. For example, confirmation bias, or selective exposure \cite{Bakshy2015,Cinelli2020} in the case of controversial topics such as politics or vaccines, are the causes of polarization and echo chambers \cite{flaxman2016,vicario2016,Cookson2020,cinelli2021,DiMarco2022}.

Besides the debate on controversial topics, our attention is also driven by our interests, which reflect our personality and culture~\cite{Larson2002, MOUNT2005,Wu2007,Hofstede2004}. In the information foraging process~\cite{Pirolli1999,Pirolli1995}, platforms may influence us; they use recommendation algorithms and other features~\cite{schmidt2017anatomy,cinelli2020covid,bakshy2015exposure} to suggest new content and potentially shape our interests~\cite{Sintas2015}.

Before the advent of personalized recommendation platforms, users organized themselves into topical groups that still exist and can take various forms: pages, channels, lists, and communities of interest. The latter form seems especially relevant to capture the dynamics of users’ interests online and is the basis of several social media platforms (and originally online fora) such as Reddit.

In such a framework, where users consume information and content in an interest-driven manner, we aim to investigate the following research questions:

\begin{description}
    \item[{\bf RQ1}] Do users tend to explore large portions of the digital space (i.e. social media platform) that they use for content consumption? 
    \item[{\bf RQ2}] Do users tend to be volatile, i.e. characterised by continuous shifts over time, with respect to the communities they interact with?
    \item[{\bf RQ3}] Do users tend to be active within communities concerning the same topics over time?
    % \item[{\bf RQ4}] \mc{THIS RESEARCH QUESTION IS ACTUALLY MATTER FOR DISCUSSION/IMPLICATIONS WE CANNOT REALLY PROVIDE AN ANSWER} Do different platforms that share a similar design but different moderation policies show distinct behaviours? 
\end{description}

 This paper addresses these research questions by analyzing two datasets of comments from Reddit and Voat (a former alternative to Reddit with no content moderation policy) in 2019. Our data includes $\approx2B$ comments from Reddit and $\approx 2M$ comments from Voat. Both platforms have an akin concept of community: Subreddits and Subverses, respectively. We present our results from both a user-centric and a community-centric perspective. This allows us to cross-check our findings using consistent data with different levels of detail.

% \mc{DID WE INCLUDE THE RESULTS ABOUT THE EMBEDDINGS IN THIS PART OF THE DISCUSSION?}

From the users' perspective, our analysis reveals that the number of communities explored by users during the year grows sublinearly in time, indicating a continuous exploration of the digital space. However, most comments are concentrated in a small set of communities. 
We operationalise the set of preferred communities and we find that its size stays constant during the observed time span, reflecting the limited attention of users. Furthermore, by defining a vector space that represents communities’ topics, we also find that, even though the preferred communities of users change over time, the topics they interact with remain roughly similar.

From the communities’ point of view, we find a high turnover of users, as shown by the comments distribution of each group and by Correspondence Analysis. To describe this framework, we propose a mathematical expression that captures the similarity between commenting users in a community during each month of $2019$. Interestingly, the similarity follows a power-law distribution, which implies that most users leave quickly, but a few stays for a long time. Finally, using Information Theory tools, we quantify the level of volatility that a community can achieve.

Our results show that users’ interests and attention are influenced by both their personal preferences and the platforms’ features. Users explore new communities over time but tend to focus on a few topics and (a small set of) communities that change over time and match their interests. Communities experience a high turnover of users, with only a few loyal ones. This implies that communities need to balance attracting new users and retaining existing ones and that platforms need to consider the trade-offs between diversity and relevance in their recommendations.

The paper is structured as follows. In the second section, we present some related works. In the third, we explain how we collect the data and how we decide to filter the initial dataset to avoid noisy results. Then, in the fourth section, we present our analysis.

\section{Related works}

Many works have focused on different aspects of Reddit and Voat. In particular, related to the users' behaviour, some studies have highlighted how it's possible to predict the involvement in conspiracy communities by examining their interaction network and their language \cite{Klein2019,Phadke2021}. In these cases, the results show that dyadic interactions with members of conspiracy groups are the most important precursor to joining these communities. Other efforts have been made to understand the migration of users \cite{DaviesAshfordEspinosa-AnkePreeceTurnerWhitakerSrivatsaFelmlee2021} using macro and micro scale analysis that reveal sub-network structures that indicate overlapping of user communities and consequences of platform moderation. In particular, also the special case after a ban of Subreddits has been studied \cite{Chandrasekharan2017,HortaRibeiro2021,Newell2021}. The results of these works seem to suggest that bans have been useful to reduce the number of active users and the usage of toxic language. 
Other studies focused on how migration happens between conflicting communities \cite{Datta2019} and in \cite{DaviesAshfordEspinosa-AnkePreeceTurnerWhitakerSrivatsaFelmlee2021} authors show a relation between migration and controversy of the topics covered in a certain group. 

Some efforts have been done also to understand the flow of users' attention between subreddits, using Graph Theory tools \cite{Rollo2022}. Finally, some research studies the concept of loyalty in Subreddits and, in \cite{tan2015all,hamilton2017loyalty}, 
 the authors find that users' propensity to become loyal is apparent from certain patterns of their first interaction with a community. 

\section{Materials and Methods}

In this section we introduce two data sources, namely {\it Reddit} and {\it Voat}. In particular, we explain how we obtained the data and the preprocessing phase that lead to the final dataset.

\subsection{Data collection}
{\bf Reddit:} Reddit is a social content aggregation website, organized in communities constructed around a certain topic, named {\it subreddits}. Each user has an account corresponding to a user name used to post submissions or to comment on other submissions and other comments. In addition, users can also upvote or downvote a submission in order to show their appreciation or criticism for it. Differently from other social media, Reddit's homepage is organized around subreddits and not on user-to-user relationships. Therefore, subreddits chosen by users are likely to represent their preferred topics and the main source of information consumed on the website. We collect public data from Reddit using the Pushift collection \cite{Baumgartner2020}. 

\noindent {\bf Voat:} Voat.co was a news aggregator website, open until 25 December 2020.
It has become famous as a place of migration for communities banned from Reddit \cite{Mekacher2022}. Its form was very similar to Reddit: discussions occurred in specific groups of interests called {\it subverses}. Users could subscribe to these communities and interact using comments, upvotes and downvotes, in a similar fashion to Reddit. In this paper, we used the dataset collected in \cite{Mekacher2022}.

\subsection{Preprocessing}
We start with all the comments published on Reddit and Voat in $2019$. Initially, we removed 8Chan, Anon and QRV because they provide identity anonymization to their users, i.e. users post with a code different from their usual username. For each community, we define its {\it size} as the number of unique users that comments on it through $2019$. Then, we select only subreddits and subverses that have a size of at least $30$ users and that are commented on at least once in each month of $2019$. 

% Finally, our data contains for each user the number of comments he left in a certain community for each month.
Table \ref{tab:breakdown} shows a data breakdown of our final datasets.

\begin{table}[!ht]
    \centering
    \begin{tabular}{|c|c|c|c|}
    \hline
         & Users & Comments & Subverses  \\
         \hline
        Voat &  23189 & 2206665 & 237 \\
        \hline
    \end{tabular}\\
    \vskip 5mm
    \begin{tabular}{|c|c|c|c|}
    \hline
         & Users & Comments & Subreddits  \\
         \hline
        Reddit & 16458011 & 215496434 & 24780 \\
        \hline
    \end{tabular}
    \caption{Data breakdown of our dataset after the preprocessing phase.}
    \label{tab:breakdown}
\end{table}

\subsection{Embeddings}\label{sec:embedding}
% \mc{ADD REFS}
We utilized the all-mpnet-base-v2 model to generate embedding vectors for each subreddit and subverse, using their respective names and descriptions as input. The all-mpnet-base-v2 model is a sentence-transformer model that encodes short paragraphs into a 768-dimensional vector space, capturing semantic information \cite{mpnet}. This model has been shown to outperform other sentence transformers based on evaluations done on 14 different datasets that measure the quality of embedded sentences \cite{model_benchmark}. Prior to generating vectors, we conducted text preprocessing by splitting CamelCase naming formats, removing special characters, numbers, punctuation, and converting the text to lowercase.

\section{Results and Discussions}

\subsection{Users point of view}\label{sec:users}
\begin{figure}[!ht]
    \centering
    \includegraphics[width=\linewidth]{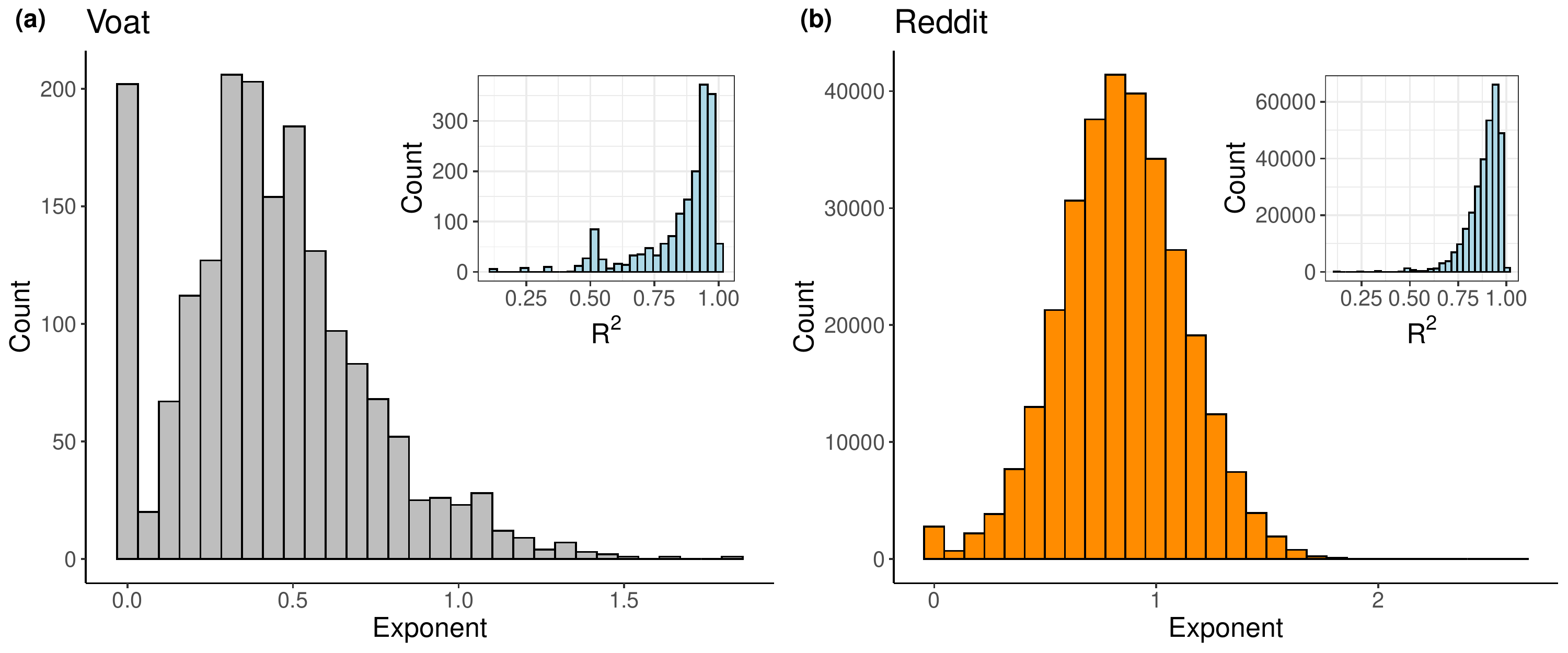}
    \caption{Distribution of exponents for $(a)$ Voat and $(b)$ Reddit.
    The insets show the distribution of $R^2$.}
    \label{fig:users_exploration}
\end{figure}

A central question for the understanding of users' behaviour online regards their exploration of new communities. We denote with $C_i (t)$ the number of known communities of user $i$ at time $t$, i.e. communities in which $i$ commented at least once. For each user with at least 50 comments, we compute $C_i (t)$ throughout the year. Interestingly, we find a time series that can be well approximated by a function such as $C_i (t) \approx t^\alpha_i$. We then apply a linear regression on the logged version of the data to obtain the exponent value and study its distribution.
Figure \ref{fig:users_exploration} shows the results for Voat (a) and Reddit (b). 

For both social media, the exponent distribution of the whole set is homogeneous (i.e. it resembles a Gaussian distribution) across the population studied, with $\alpha_i$ peaked at $\Bar{\alpha} = 0.43$ for Voat and $\Bar{\alpha} = 0.85$ for Reddit. 
The insets show the $R^2$ distribution of fittings, indicating that for the majority of them, explanatory variables cover the greatest part of the variance in the data.

It is worth noticing that, in the case of Reddit, there is a lower percentage of users for which the fitting returns an exponent equal to $0$, meaning that no new communities are discovered during the year. This effect is less prominent on Voat probably because of the nature of the platform, which hosts communities banned by Reddit~\cite{Mekacher2022, Newell2021}. In fact, it is possible that users subscribe to Voat only to join the same community banned on Reddit, with no further interest for others~\cite{deplatforming2023}.
What is displayed in Figure~\ref{fig:users_exploration} indicates that users tend to explore new communities following a sublinear behaviour, confirmed by the values of $\alpha$ mostly below one. This first analysis outlines the tendency of users to explore just a small portion of the available space but does not provide indications about how they allocate their attention/activity in this digital environment.

To investigate this aspect, we define $PC_i(t) = \{c_1,\ldots,c_n\}$ as the set of communities preferred by user $i$ at time $t$. In particular, we say that a community $c_j \in PC_i (t)$ if user $i$ allocates at least $20$ comments in $c_j$ at month $t$. This threshold is justified by a power-law behaviour of the users in terms of community commenting due to which a value of 20 comments can be considered a relatively high one (approximately $70\%$ of users left fewer comments than this threshold).

\begin{figure}[!ht]
    \centering
    \includegraphics[width = \linewidth]{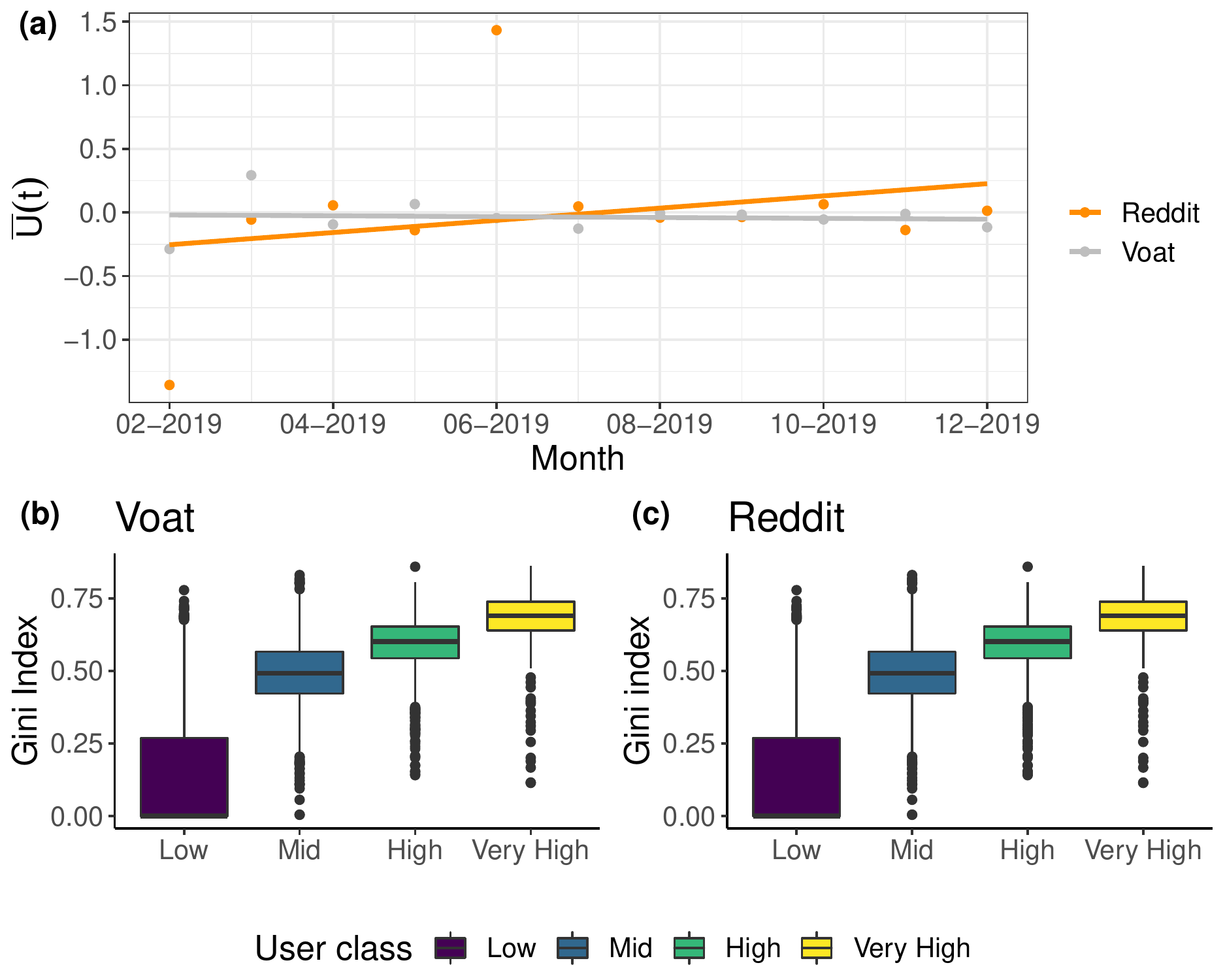}
    \caption{$(a)$ fitting of $\bar{U}$ for Voat and Reddit. $U_i(t)$ is an unbounded quantity thus observing $\bar{U}$ ranging around the interval [-1,1] already signals a tendency of the users to interact with a relatively stable number of communities.
    $(b)$ and $(c)$ show Gini Index computed for each user on Voat and Reddit.}
    \label{fig:pc}
\end{figure}

We are interested in characterising how this set evolves over time for each user. In particular, we define $A_i (t)$ as the number of communities added to $PC_i(t)$ and with $D_i (t)$ the number of communities removed from $PC_i (t-1)$. We compute for each month the difference $U_i (t) = A_i (t) - D_i (t)$ for all the users that have at least one preferred community per month and we denote with $\bar{U}(t)$ its mean value. Then, we employ a linear regression model $\Bar{U}(t) = mt + q$ to understand how $\bar{U}(t)$ evolves over the year. Figure~\ref{fig:pc} shows the results. 

For both Voat and Reddit, we find that the slope is not significantly different from $0$ ($p = 0.82$ for Voat and $p = 0.45$ for Reddit). 

This result, joined with the previous one, provides a response to {\bf RQ1} by corroborating the hypothesis that users have limited  attention and tend to interact with a fixed number of preferred communities and information sources. Interestingly, if we interpret a community as a point of interest in the virtual space, our result resonates with some findings about the limitations observed during the exploration of the physical space for human mobility~\cite{Alessandretti2018,Song2010,Isaacman2011,abramo2017investigation,de2019strategies}.

To deepen the investigation of users' attention and provide a different point of view, we divide them into classes of activity. In particular, for a fixed community, we assign to each user a class indicating its activity in it. To define classes we use a non-parametric method developed for partitioning heavy-tailed distributions developed by~\cite{citation_impact} and recently employed in different domains~\cite{abramo2017investigation,cinelli2021ambiguity}. Briefly, we compute the mean number of comments $\Bar{x}$ and users that have left less than $\Bar{x}$ are assigned to class {\it low}. Then we delete these users from the distribution and recursively repeat the procedure using four classes of activity $\{low,mid,high,very \; high\}$.

Figure \ref{fig:pc} shows the distribution of the Gini index~\cite{gini1912variabilita}, representing the degree of inequality of a distribution per user class. %It represents the degree of inequality of a distribution. A value near $1$ implies that comments are distributed mostly in only one community while a value near $0$ indicated that those comments are distributed equally across communities.
For both Social Networks, users in higher activity classes tend to have a higher Gini Index. Therefore we find that higher volumes of interactions are less likely to be distributed homogeneously across communities.

% With the analysis presented so far, we cannot deduce if these communities remain the same or change over time thus we cannot still address RQs 2 and 3.

\subsection{Communities point of view}

In the previous section we showed that users have a limited attention span and that they tend to interact with a fixed (small) number of preferred communities. 
In this section, our focus shifts to how this heterogeneity manifests from the communities' point of view. Examining the users' exploration process in the digital space considering the communities perspective, may offer further evidence in favour of what we obtained by analysing the activity from a user-centric perspective.

\begin{figure}[!ht]
\centering
\includegraphics[width=\linewidth]{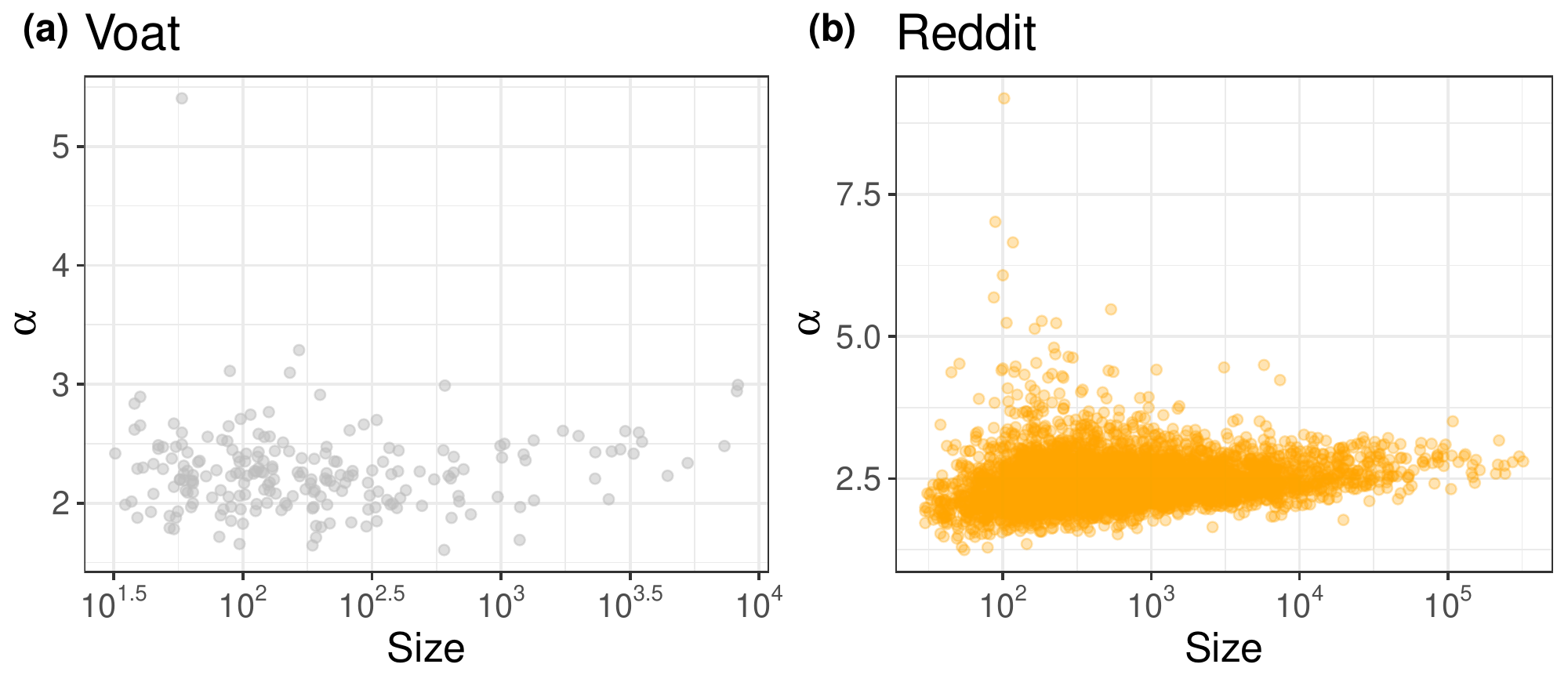}
\caption{$(a)$ distribution of $\alpha$ on Reddit compared to size of the relative community. $(b)$ shows the same but for Voat. For both $(a),(b)$ only significant fittings (i.e. $p > 0.05$) are considered). Only a sample of $5000$ points is shown for Reddit.}
\label{fig:fit}
\end{figure}

We initially consider a fixed community $c$ and the distribution $D_c$ of comments per user, i.e. the number of comments left by each user in $c$.
Using the package {\it poweRlaw} in R \cite{poweRlaw}, we fit $D_c$ with a power law distribution $p(x)\sim x^{-\alpha}$. For each community, we obtain an exponent $\alpha$ and a $p-value$ for the statistical test $H_0:$ {\it data comes from a power law distribution}. Thus, a $p-value > 0.05$ indicates that we can't reject the hypothesis that $D_c$ is a power law. In Figure \ref{fig:fit} we compare the exponents with the size of each community.

The results show that for Reddit the exponents tend to be $\approx 3$, while for Voat they tend to be $\approx 2.5$, meaning that a minority of users contribute to most of the comments within communities. Although the $p-$values distribution is right-skewed, for most of the communities it is not possible to reject $H_0$ at a significance level of $0.05$ (see Figure~\ref{fig:p_value_power_law} in SI). In more detail, approximately $72\%$ of subreddits have $p > 0.05$ and approximately $89\%$ of subverses satisfy the same property. Therefore, most users left the community early, while only a minority stay longer, even if no information is obtained about their time of permanence. 

In Figure~\ref{fig:turnover} and Figure~\ref{fig:heatmap} of SI we provide anecdotal evidence, deriving from an exploratory analysis regarding the users' activity, that the (few) preferred communities of users are constantly changing. In fact, consistently with previous works~\cite{tan2015all,hamilton2017loyalty} but from a community point of view, we observe that the great majority of users leave communities quickly and that their engagement is one of the determining factors for remaining. 
Along with the results presented in the previous section, there is evidence that users maintain a stable, fixed number of communities, but these communities are constantly changing.  

\begin{figure}[!ht]
\centering
\includegraphics[width = \linewidth]{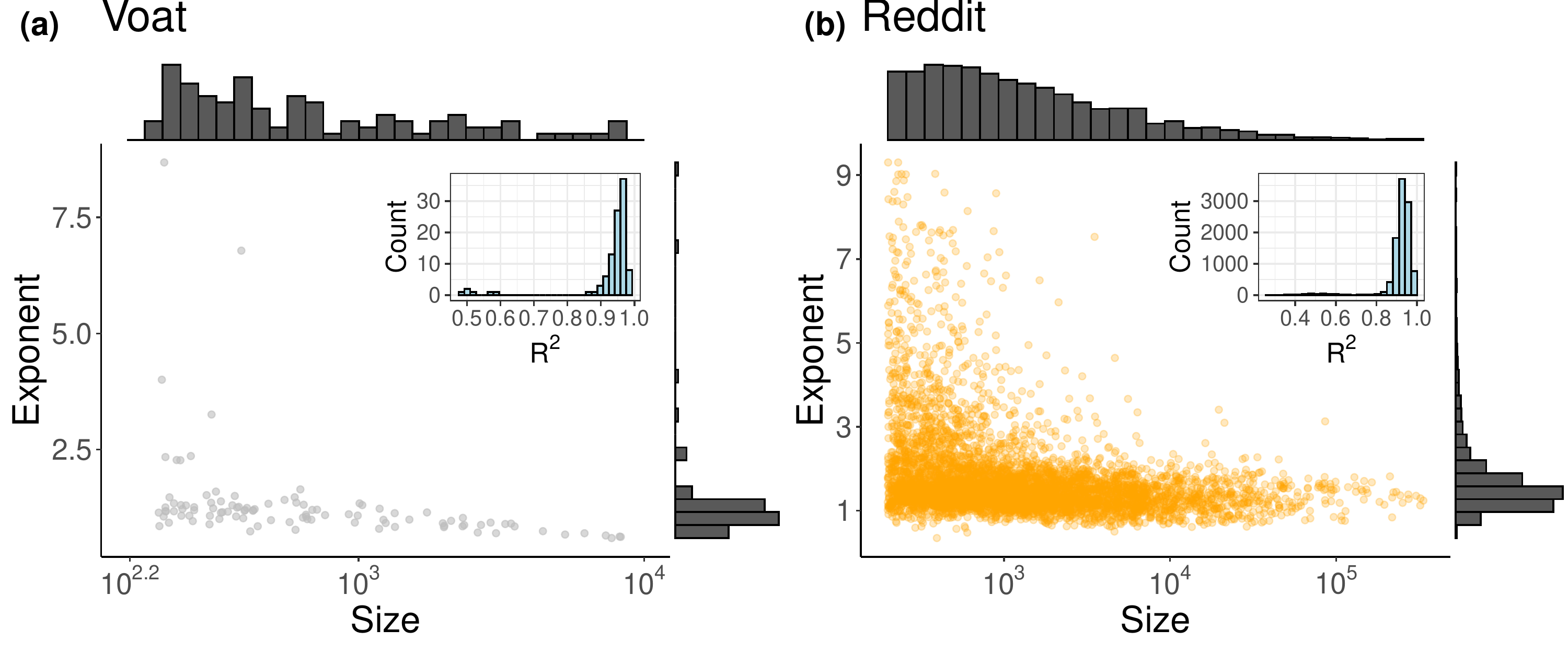}
\caption{$(a),(b)$ distribution of exponents' values for Subreddits and Subverses. The insets show histograms of $R^2$. Only a sample of $5000$ points is shown for Reddit.}
\label{fig:heatmap_fit}
\end{figure}

To generalise and extend the results suggested by Figure~\ref{fig:turnover} and Figure~\ref{fig:heatmap} of SI, we compare the number of users that remain in each community considering January as a reference month. Namely, we compare the set of users commenting in January with those commenting in the other eleven months with {\it Jaccard Index}, which can be used to measure the similarity between the set of users that comment in different pairs of months.
In particular,

\begin{equation*}\label{eq:Jaccard}
    J(t) = \frac{|\{user \; in \; January\} \cap \{user \; in \;t \}|}{|\{user \; in \; January\} \cup \{user \; in \; t\}|}
\end{equation*}

where $t$ is the month being considered.
The obtained values suggest that the number of common users decays in time for both Voat and Reddit with a behaviour similar to a power law. The exponent of this power law would then characterize how long users give attention to a certain community.
Figure \ref{fig:heatmap_fit} shows the results of the fittings and compares $\alpha$ with the size of the community. In particular, we perform a linear regression on the logged version of the data and, to avoid noisy inputs, we employ only subverses of size greater than 200 and subreddits of size greater than $10^3$.  

We note that subreddits with a greater size tend to have a value of the exponent approximately $1$, while in Voat they seem to decrease with the size of the subverses. The insets show the distributions of $R^2$, which confirm that the majority of the fittings explain most of the variance in the data. 
The results of Figure~\ref{fig:heatmap_fit}, regarding the similarity of the users set indicate that for most subreddits, the decay is superlinear, i.e. users populating them are changing fast, with the exception of the biggest ones that seem to attract users for a longer time. On the other hand, most of the subverses have a sublinear behaviour, i.e. users tend to stay longer in the same communities. It is worth noticing that, for this last case, the bigger the subverse (in terms of size), the lower the exponent, meaning that (as in Reddit) big subverses tend to attract users for a longer time.

We are now interested in giving another characterization of the users turnover from a community point of view. In this context, we use Correspondence Analysis (CA) that, compared to the previous analyses, is able to highlight the patterns realized by users on each community over time.
Fixed a community $k$ we consider the contingency table $B_k$ of dimension $n \times 12$, where $n$ is the number of unique users that have commented in $k$. The entry $B_k (i,j)$ can be either $0$ or $1$ depending if the user $i$ comments $k$ on month $j$.
CA is used for visualizing rows and columns of $B$ as points in a low-dimensional space (two in our case), such that the positions of the row and column points are consistent with their associations in the table. Since we are interested in turnover, we plot only the points relative to months.

\begin{figure}[!ht]
\centering
\includegraphics[width=\linewidth]{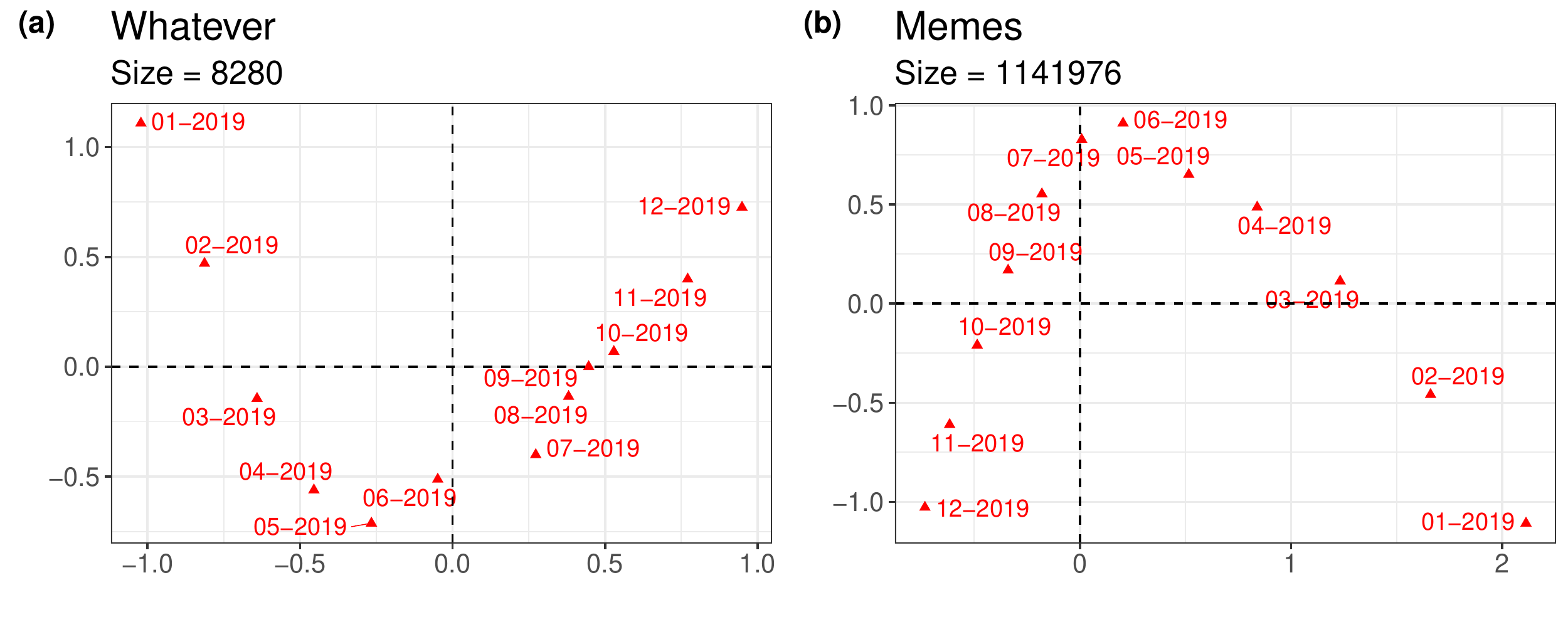}
\caption{Results of $CA$ applied to a subverse $(a)$ and a subreddit $(b)$.}
\label{fig:ca}
\end{figure}

We apply CA using {\it FactoMineR} package in R \cite{CA} to $B$.
Partial results are shown in Figure \ref{fig:ca} (see SI for the coordinate obtained for subverses of size at least 1500 and subreddits of size at least $10^5$).

Panel $(a)$ shows the $CA$ applied to {\it Whatever}, a popular subverse. The quadratic shape indicates that different months tend to be commented on by different users. In particular, distant months (in time) have approximately a bigger distance, indicating that users have a limited lifetime in the same community. Otherwise, if the same set of users would comment constantly on the same community we would obtain a higher concentration of points in the two-dimensional space.
Panel $(b)$ shows the results of $CA$ applied to {\it Memes}, a famous subreddit. A similar shape is shown, indicating a high turnover of users. In general, $CA$ provides a technique to highlight turnover in communities as shown in the tables presented in SI.
%In particular, the turnover is much more highlighted in communities with high sizes. 

Finally, we can confirm what obtained from Section~\ref{sec:users} from a community perspective.

\subsection{Volatility over communities}

% \mc{Let's make a small preamble to state that we provide an answer to RQ2 and what we do here and what is volatility (see how I modified RQ2 in the introduction)}

In this section we aim to answer to {\bf RQ2}. In particular, we use Information Theory tools to provide a mathematical formulation of how much users are volatile from a community point of view.

Consider a community $k$ and a user $i$, we count the number of months $i$ comments $k$ and we denote this quantity as $m(i,k)$. Obviously, $1 \leq m(i,k) \leq 12$. We consider the distribution of $m(i,k)$ for each community $k$. 

We try to measure volatility using the following idea: if a community is composed of loyal users, then the distribution $m(i,k)$ should be something similar to a uniform distribution, i.e. the users are constantly active in the community. On the other hand, if a community is largely composed of volatile users, then $m(i,k)$ should be concentrated in one point, like a delta distribution. These two behaviours can be captured by the concept of (normalized) Shannon Entropy $H(x)$. In particular, a community of non-volatile users should show an entropy near $1$, while a volatile community should have a value near $0$. Since $H(x)$ can be interpreted, in this context, as a measure of stability, we summarise the volatility of users within a community using the complement of Shannon's entropy that we indicate as $V(x) = 1 - H(x)$, where $x$ is the community under consideration.

\begin{figure}[!ht]
\centering
\includegraphics[width=\linewidth]{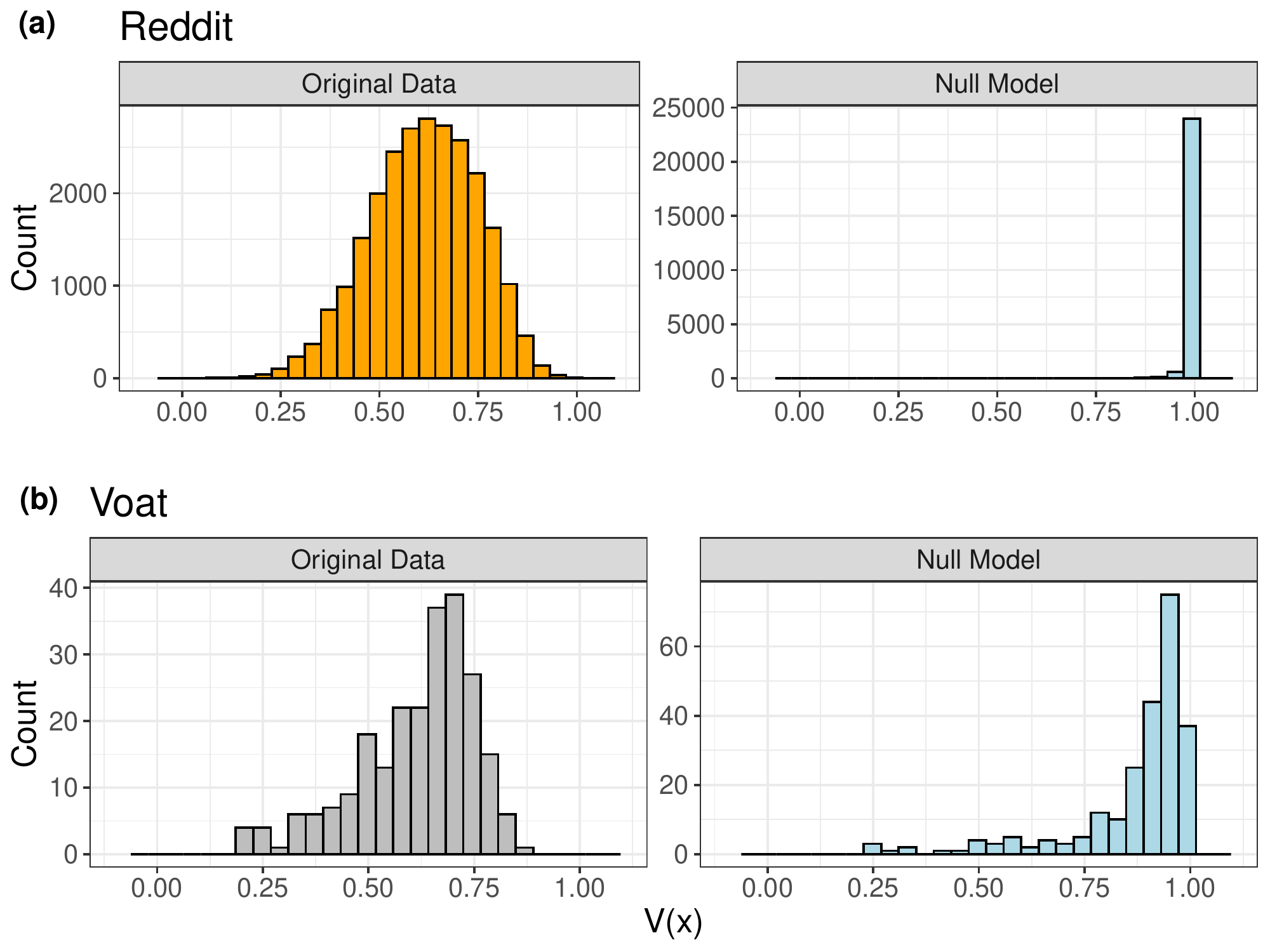}
\caption{Comparison of $V(x)$ distributions for real data and a random null model. $(a)$ is obtained with Reddit Data while
$(b)$ is obtained with Voat Data.}
\label{fig:loyalty}
\end{figure}

Figure \ref{fig:loyalty} shows the distributions of $V(x)$ for each subreddit and subverse. 

% \mc{There is a typo here that I couldn't fix...}
Panel $(a)$ of Figure~\ref{fig:loyalty} displays the distribution of the volatility for each subreddit. On the right, it is shown the distribution of volatility for a null model in which the interactions between users and communities are randomised. Panel $(b)$ shows the same analysis on Voat.

It is worthwhile to notice that the volatility distribution has a left-skewed shape for Voat, while it looks more centred in the case of Reddit. This can be due to the low volume size effects of Voat, compared to Reddit. 

It's also interesting to note that both random models show very left-skewed distributions, essentially different from the original ones. This suggests that in a random model users show higher volatility in the communities in which they comment. Moreover, their behaviour follows a non-random pattern and, despite being heterogeneous, their activity may be driven by some factors that we aim to investigate in the next section.

\subsection{Evolution of user interest}

In the previous section we have shown evidence that users continuously explore new communities, maintaining a small set of preferred ones. 
Here, we deepen the analysis by studying how their interests evolve over time considering a semantic representation of each community. In other words, we aim to understand if the observed volatility implies also a shift in the set of users' interests or if they tend to make small variations. For the analysis, we embed each community in a vectorial space in which the coordinates indicate the topics discussed in that (see Section~\ref{sec:embedding} for further details).

We consider the same set of users considered in Section~\ref{sec:users} and their preferred communities. Let $u$ be a user and $PC_u (t)$ the set of $u$'s preferred communities at month $t$. Moreover, denote with ${\bf v_c}$ the embedding vector of community $c$. We define the interest $s_u (t)$ of users $u$ at month $t$ as the centroid of the vectors associated with $u$'s preferred communities, namely:
\begin{equation}
    s_u (t) = \sum_{i \in PC_u (t)} \frac{{\bf v}_i}{768} .
\end{equation}
After computing these quantities, we compare them between consecutive pairs of months for each user using cosine similarity. We employ a linear regression model to fit their evolution throughout the year.
Therefore, for each user, we obtain the values of the slope and intercept of the fitting together with their p-values.

\begin{figure}[!ht]
    \centering
    \includegraphics[width = \linewidth]{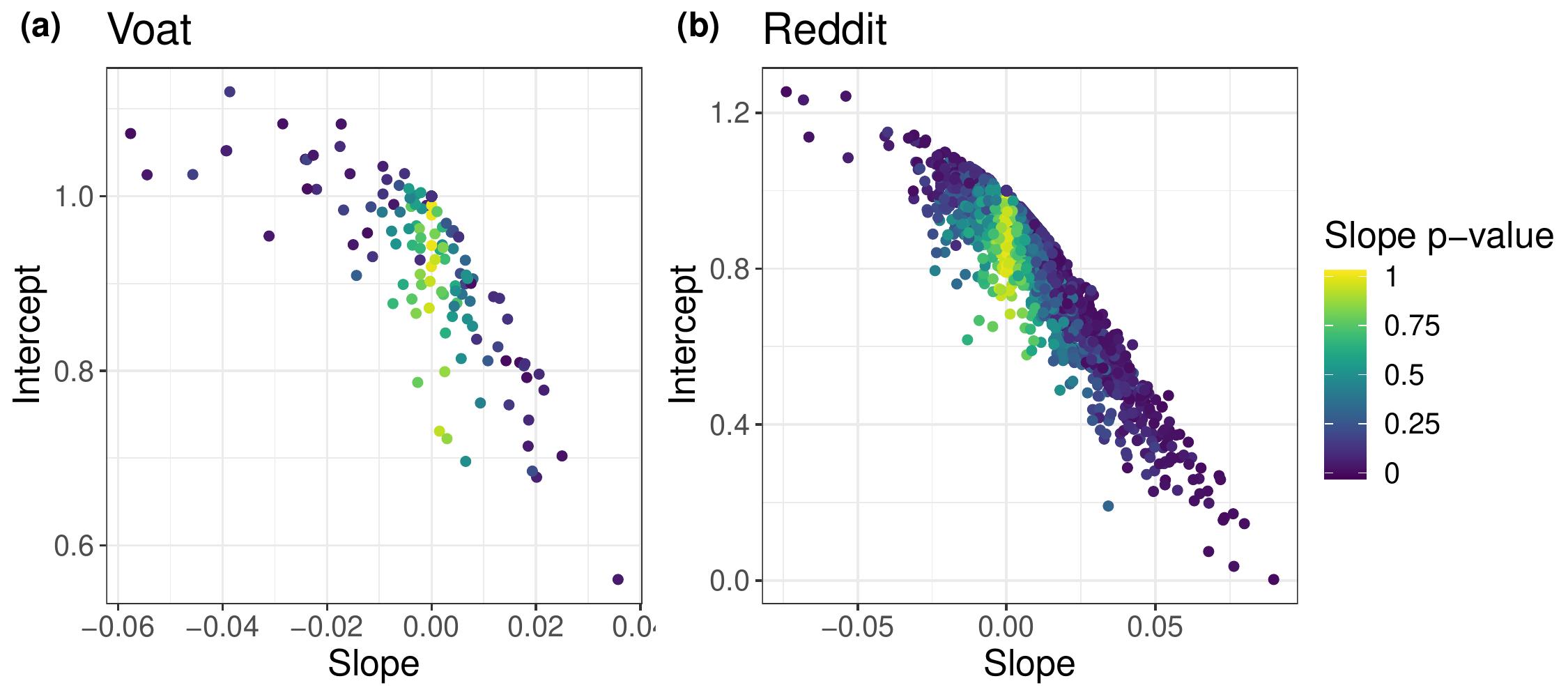}
    \caption{Bivarite distribution of the slope and intercepts obtained using a linear regression model to fit the evolution of cosine similarity of topics throughout the year. For representation purposes, only a sample of 2000 points is shown for Reddit.} 
    \label{fig:mean_interest}
\end{figure}

Figure~\ref{fig:mean_interest} shows the bivariate distributions of the slope and intercept, colored using the $p-$value associated with the former, for each social media.

The results are similar for both Voat and Reddit. In particular, a high percentage of fitting in which the cosine similarity is constant throughout the year is present, with relatively high values of the intercept. Moreover, when this is not the case, the values of the slope are not significantly different from zero, indicating approximately constant (high) values of cosine similarity. 

The analysis, considering also previous sections' results, corroborate the idea that, despite the continuous exploration of new spaces,  users tend to visit communities that are about the same topics mostly. Thus, we can provide a positive answer to {\bf RQ3}.

\section{Conclusions} 
In this paper we analyze the behaviour of a large dataset of users populating communities on Reddit and Voat. 
First of all, we find a negative answer to {\bf RQ1}. In fact, users tend to continually explore new communities, but with a sublinear behaviour. However, despite the new communities discovered, they tend to stay only in a small subset of them, constantly updated throughout the year.

Instead, we obtain a positive answer to {\bf RQ2}. In fact, both regressions and CA indicate that the majority of users tend to leave very soon the communities. Moreover, using information theory tools, we also find that users tend to follow a behaviour that cannot be reproduced by a random null model.

Using embedding vectors of the communities space we also find evidence for a positive answer of {\bf RQ3}, since the communities explored by users tend to cover essentialy the same topics.

% We find that the limited attention spans of users reflect a high turnover of them from the communities point of view. In particular, we find that Correspondence Analysis is able to detect the changing of users through time and information theory tools can be successfully used to define pragmatically the concept of loyalty in a community.

It is worth noticing that all the obtained results are very similar for Voat and Reddit. In particular, it seems that, although Voat was born as a place of migration for communities banned from Reddit (mostly questionable ones), this does not play a role in users' behaviour and, instead, only the similarities in their organization influence that.

Finally, our results suggest that interests lead the exploration of new communities but the necessity of social feedback also leads users to stay in those communities that are more active, despite the presence of moderation policies.

\bibliographystyle{unsrt}
\bibliography{sample}

\end{document}

\maketitle
\section{Fittings}
\begin{figure}[!ht]
    \centering
    \includegraphics[width = \linewidth]{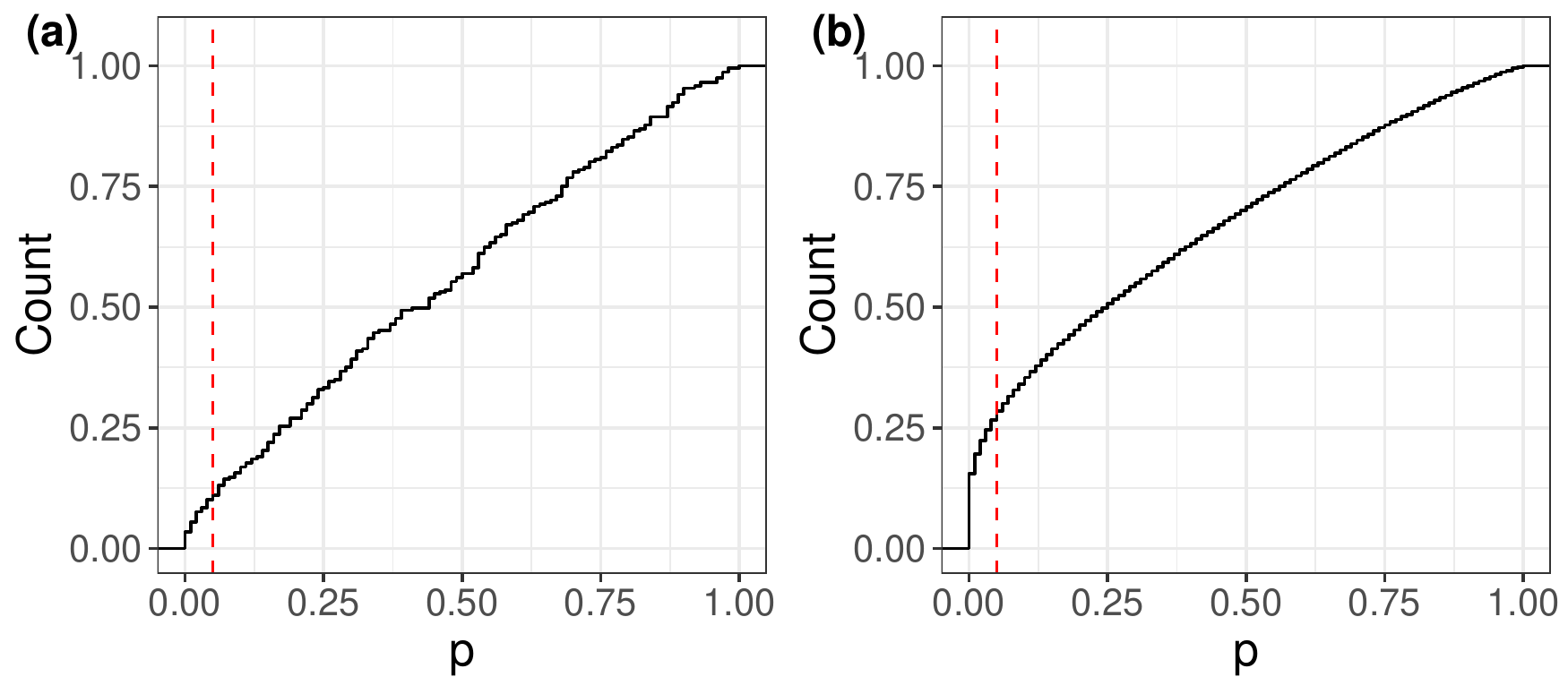}
    \caption{$p-$values for the fitting of distribution of comments in each $(a)$ subverse or $(b)$ subreddit. The line $x = 0.05$ is highlighted in red.}
    \label{fig:p_value_power_law}
\end{figure}

Figure~\ref{fig:p_value_power_law} shows the cumulative distribution of the p-values for the fitting of the distribution of comments $D_c$ of each community.

The red line $x = 0.05$ is highlighted to show that, for the majority of the fitting, we cannot reject the hypothesis that $D_c$ comes from a power law.

\section{Turnover of users}
\begin{figure}[!ht]
\centering
\includegraphics[width=\linewidth]{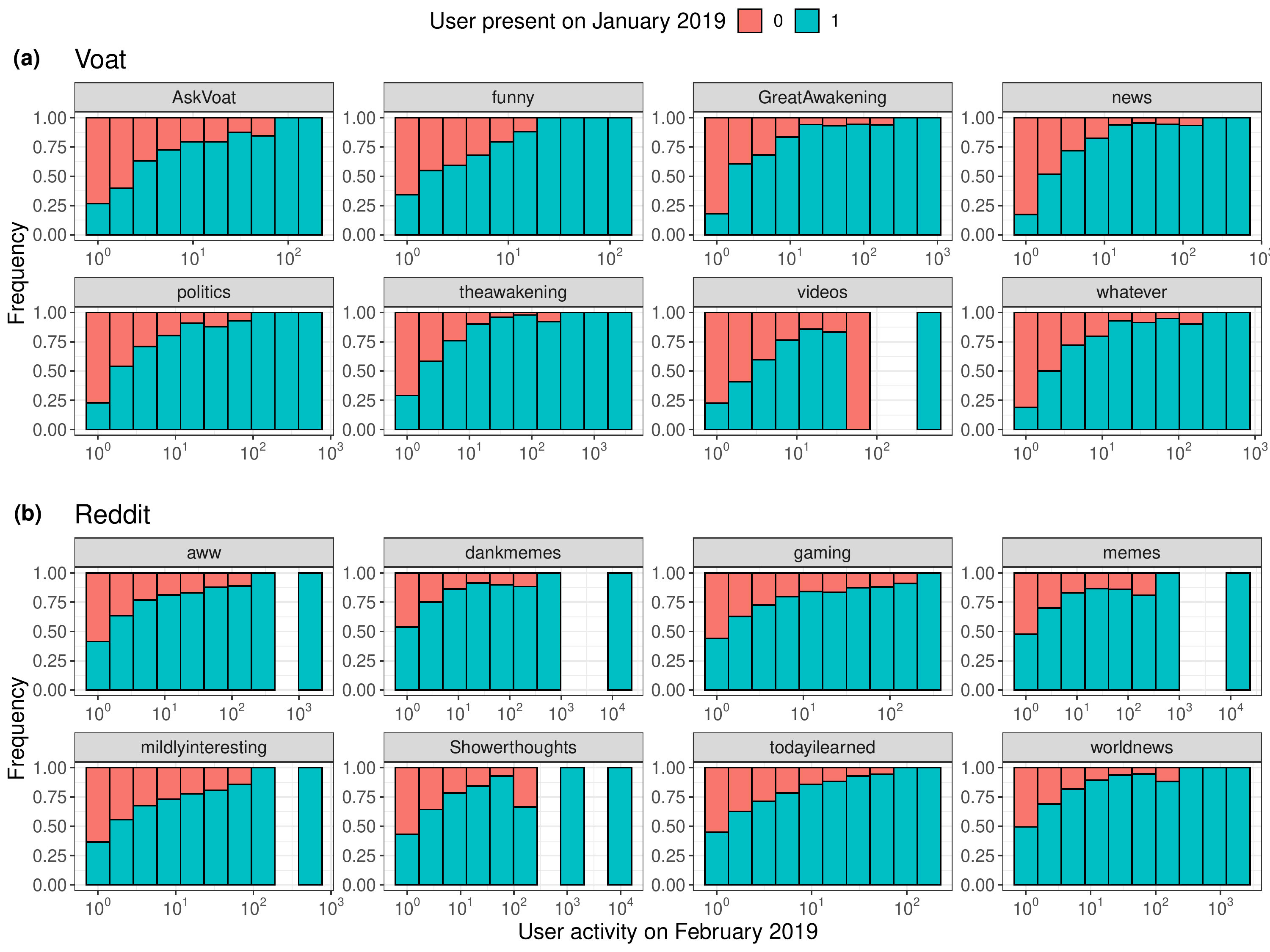}
\caption{$(a)$ histogram of activities of Voat users in the biggest 9 subverses on February 2019. $(b)$ shows the same on the biggest 9 subreddits of Reddit. The blue colour highlights users that were present in the same community also in January 2019.}
\label{fig:turnover}
\end{figure}

To investigate the turnover of users, we do an exploratory analysis regarding users' activity. In particular, Figure \ref{fig:turnover} compare the activity of user between January and February $2019$ in the $9$ biggest community of Voat and Reddit. The blue colour in each bar shows the percentage of users that are present in both months.

We see that users with higher activities tend to return to their communities. In particular, this effect occurs in both Reddit and Voat. This particular effect suggests that activity is an important benchmark to understand how long a user will stay in a certain group.

Recall that Jaccard similarity is defined as:

\begin{equation*}\label{eq:Jaccard}
    J(t) = \frac{|\{user \; in \; January\} \cap \{user \; in \;t \}|}{|\{user \; in \; January\} \cup \{user \; in \; t\}|}
\end{equation*}

\begin{figure}[!ht]
\centering
\includegraphics[width = \linewidth]{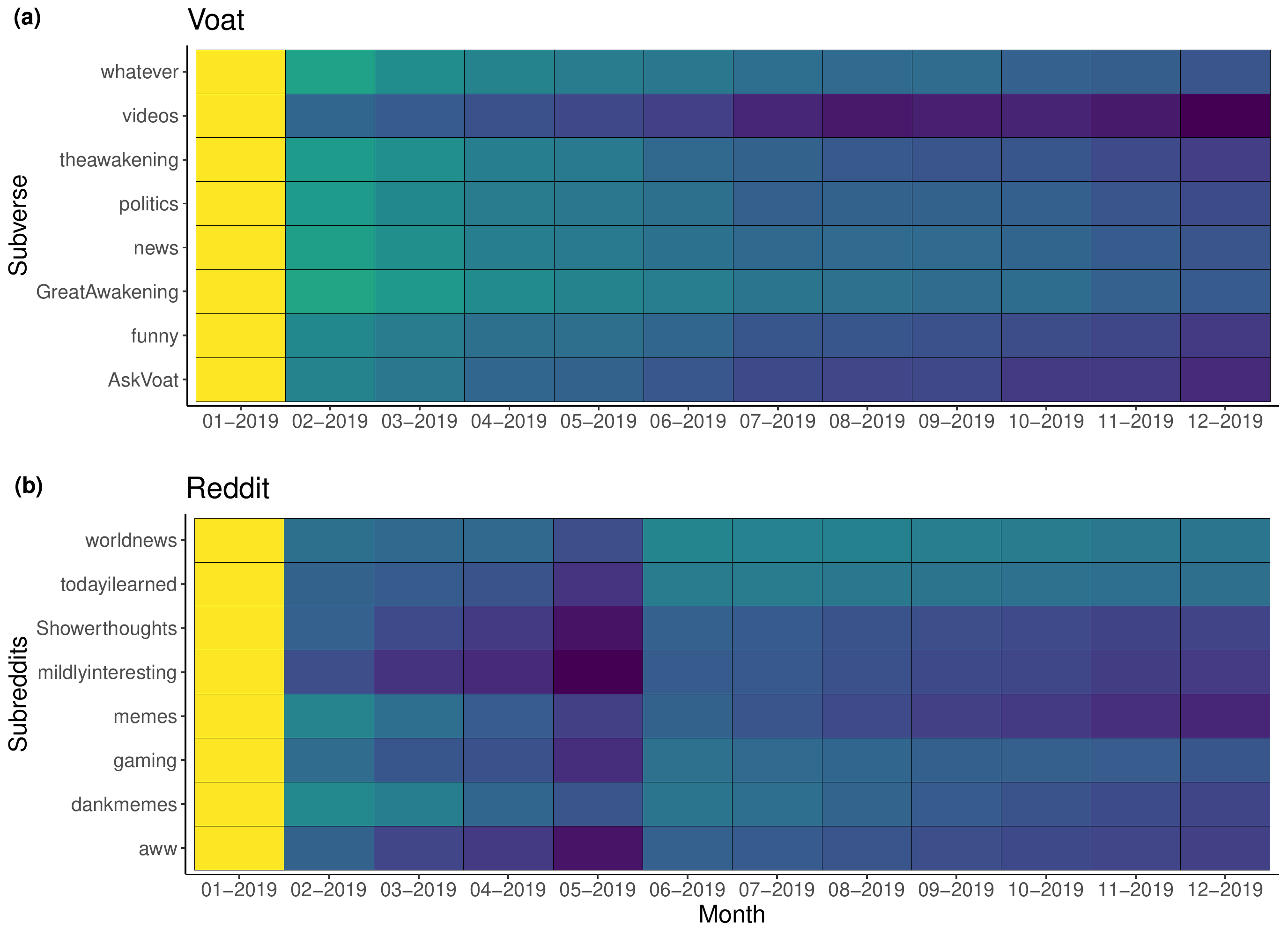}
\caption{These heatmaps show the {\it Jaccard Index} $J(t)$ between January and a specific month for the 8 biggest communities in $(a)$ Voat and $(b)$ Reddit. The colour shows the intensity of $J$ using a logarithmic scale.}
\label{fig:heatmap}
\end{figure}

We use \eqref{eq:Jaccard} to compare the similarity of users commenting in different months of 2019. Figure~\ref{fig:heatmap} shows two heatmaps in which each point is filled with the relative value of \eqref{eq:Jaccard}.

The values suggest a power-law decay for the similarity, indicating that most users left early.

\section{Results of Correspondence Analysis}

In this section we show the coordinates obtained through CA applied to Subverses (in Table~\ref{tab:CA_voat} and Subreddit (in Table~\ref{tab:CA_reddit}). To avoid noise signals from small communities we only report results from subverse with a size of at least $1500$ and subreddit with a size of at least $10^5$.

\begin{table}[!ht]
\hspace{-26.5mm}
\scalebox{0.65}{
\begin{tabular}{rlllllllllllll}

  \hline
 & Subverse & 01-2019 & 02-2019 & 03-2019 & 04-2019 & 05-2019 & 06-2019 & 07-2019 & 08-2019 & 09-2019 & 10-2019 & 11-2019 & 12-2019 \\ 
  \hline
1 & whatever & (-1.02, 1.11) & (-0.81, 0.47) & (-0.64,-0.14) & (-0.45,-0.56) & (-0.27,-0.71) & (-0.05,-0.51) & ( 0.27,-0.40) & ( 0.38,-0.14) & ( 0.45, 0.00) & ( 0.53, 0.07) & ( 0.77, 0.40) & ( 0.95, 0.73) \\ 
  2 & news & (-1.15, 0.92) & (-0.79, 0.30) & (-0.62,-0.11) & (-0.40,-0.55) & (-0.16,-0.62) & ( 0.02,-0.67) & ( 0.29,-0.58) & ( 0.37,-0.12) & ( 0.43, 0.07) & ( 0.48, 0.15) & ( 0.73, 0.53) & ( 0.87, 0.83) \\ 
  3 & GreatAwakening & ( 1.05,-0.96) & ( 0.86,-0.21) & ( 0.60, 0.22) & ( 0.34, 0.46) & ( 0.23, 0.59) & ( 0.07, 0.58) & (-0.16, 0.61) & (-0.37, 0.24) & (-0.45, 0.04) & (-0.51,-0.12) & (-0.74,-0.50) & (-0.82,-0.69) \\ 
  4 & politics & (-1.08, 0.92) & (-0.84, 0.36) & (-0.70,-0.16) & (-0.43,-0.61) & (-0.15,-0.62) & ( 0.06,-0.62) & ( 0.39,-0.58) & ( 0.44,-0.07) & ( 0.43, 0.08) & ( 0.52, 0.24) & ( 0.73, 0.59) & ( 0.85, 0.77) \\ 
  5 & funny & (-1.07, 1.21) & (-0.87, 0.38) & (-0.71,-0.21) & (-0.45,-0.62) & (-0.26,-0.63) & (-0.04,-0.53) & ( 0.24,-0.53) & ( 0.42,-0.12) & ( 0.45,-0.01) & ( 0.57, 0.12) & ( 0.79, 0.45) & ( 0.97, 0.75) \\ 
  6 & theawakening & (-1.18,-0.93) & (-0.83, 0.09) & (-0.53, 0.47) & (-0.27, 0.54) & (-0.08, 0.45) & ( 0.11, 0.56) & ( 0.29, 0.60) & ( 0.47, 0.09) & ( 0.55,-0.12) & ( 0.56,-0.30) & ( 0.72,-0.60) & ( 0.78,-0.77) \\ 
  7 & AskVoat & (-1.08, 0.94) & (-0.86, 0.38) & (-0.60,-0.13) & (-0.37,-0.48) & (-0.14,-0.64) & (-0.01,-0.76) & ( 0.35,-0.62) & ( 0.48,-0.27) & ( 0.44, 0.02) & ( 0.60, 0.18) & ( 0.89, 0.71) & ( 0.99, 0.96) \\ 
  8 & videos & (-1.15, 0.54) & (-1.10, 0.71) & (-0.61, 0.12) & (-0.43,-0.41) & (-0.14,-0.64) & ( 0.00,-0.74) & ( 0.36,-0.97) & ( 0.37,-0.24) & ( 0.55,-0.07) & ( 0.53, 0.15) & ( 0.84, 0.80) & ( 1.00, 1.02) \\ 
  9 & pics & (-1.16,-1.12) & (-0.94,-0.50) & (-0.78, 0.37) & (-0.44, 0.85) & (-0.24, 0.69) & (-0.06, 0.58) & ( 0.38, 0.65) & ( 0.51, 0.13) & ( 0.55, 0.12) & ( 0.60,-0.11) & ( 0.91,-0.68) & ( 0.88,-0.80) \\ 
  10 & technology & (-1.28, 0.89) & (-0.86, 0.32) & (-0.65,-0.25) & (-0.26,-0.77) & (-0.04,-0.98) & ( 0.20,-0.48) & ( 0.66,-0.28) & ( 0.62, 0.05) & ( 0.53, 0.23) & ( 0.49, 0.30) & ( 0.89, 1.00) & ( 0.87, 0.95) \\ 
  11 & fatpeoplehate & (-0.97, 0.73) & (-0.91, 0.60) & (-0.63, 0.14) & (-0.48,-0.24) & (-0.28,-0.62) & (-0.05,-0.54) & ( 0.25,-0.77) & ( 0.37,-0.24) & ( 0.52,-0.17) & ( 0.49,-0.06) & ( 0.84, 0.46) & ( 1.14, 1.14) \\ 
  12 & science & (-1.12, 0.90) & (-0.86, 0.29) & (-0.71,-0.02) & (-0.58,-0.60) & ( 0.00,-0.43) & ( 0.09,-0.61) & ( 0.50,-0.79) & ( 0.46,-0.36) & ( 0.59,-0.32) & ( 0.68, 0.17) & ( 0.65, 0.22) & ( 1.16, 1.55) \\ 
  13 & gaming & (-1.09, 1.04) & (-0.68, 0.45) & (-0.58,-0.03) & (-0.40,-0.67) & (-0.15,-0.59) & (-0.06,-0.50) & ( 0.40,-1.07) & ( 0.54,-0.14) & ( 0.70,-0.01) & ( 0.63, 0.22) & ( 1.08, 0.64) & ( 1.36, 1.27) \\ 
  14 & Niggers & ( 1.13,-0.85) & ( 0.99,-0.71) & ( 0.62, 0.39) & ( 0.36, 0.69) & ( 0.28, 0.94) & ( 0.11, 0.60) & (-0.32, 0.32) & (-0.41, 0.07) & (-0.54,-0.12) & (-0.54,-0.17) & (-0.67,-0.32) & (-0.84,-0.73) \\ 
  15 & milliondollarextreme & (-0.87, 0.76) & (-0.74, 0.37) & (-0.58, 0.08) & (-0.16,-0.55) & (-0.02,-0.85) & ( 0.21,-0.51) & ( 0.51,-0.44) & ( 0.70,-0.02) & ( 0.86, 0.12) & ( 0.89, 0.44) & ( 1.20, 0.84) & ( 1.32, 1.39) \\ 
  16 & movies & ( 1.23, 1.51) & ( 0.71, 0.17) & ( 0.85, 0.26) & ( 0.66,-0.41) & ( 0.60,-0.97) & ( 0.18,-0.55) & (-0.17,-1.09) & (-0.52,-0.22) & (-0.53,-0.08) & (-0.66, 0.13) & (-0.98, 0.67) & (-0.94, 0.67) \\ 
  17 & Conspiracy & (-0.98,-1.10) & (-0.85,-0.28) & (-0.76, 0.15) & (-0.37, 0.87) & (-0.11, 0.44) & (-0.04, 0.35) & ( 0.37, 0.71) & ( 0.57, 0.32) & ( 0.60, 0.12) & ( 0.68,-0.01) & ( 0.76,-0.38) & ( 1.04,-1.16) \\ 
  18 & MeanwhileOnReddit & (-1.19, 1.15) & (-0.95, 0.35) & (-0.56, 0.14) & (-0.54,-0.56) & (-0.06,-0.59) & ( 0.20,-0.67) & ( 0.48,-0.33) & ( 0.56,-0.12) & ( 0.29,-0.16) & ( 0.74, 0.01) & ( 0.87, 0.39) & ( 1.20, 1.47) \\ 
  19 & Worldnews & (-1.22, 1.30) & (-0.72,-0.09) & (-0.72,-1.31) & (-0.30,-0.49) & ( 0.05,-0.10) & ( 0.10, 0.05) & ( 0.69, 0.27) & ( 0.69, 0.08) & ( 0.67, 0.17) & ( 0.75, 0.07) & ( 0.38, 0.00) & ( 0.65, 0.30) \\ 
  20 & 4chan & (-0.96,-0.76) & (-0.91,-0.74) & (-0.61, 0.01) & (-0.25, 0.17) & (-0.30, 0.56) & (-0.13, 0.90) & ( 0.12, 1.00) & ( 0.51, 0.22) & ( 0.53,-0.07) & ( 0.61,-0.10) & ( 0.62,-0.14) & ( 1.17,-0.89) \\ 
  21 & pizzagate & (-1.56, 0.90) & (-0.55,-0.21) & (-0.40,-0.62) & (-0.27,-0.45) & (-0.15,-0.40) & (-0.05,-0.53) & ( 0.34,-0.64) & ( 0.61,-0.43) & ( 0.38,-0.11) & ( 0.56, 0.17) & ( 0.82, 1.06) & ( 0.88, 1.34) \\ 
  22 & gifs & (-1.49, 1.02) & (-0.83, 0.22) & (-0.59, 0.03) & (-0.18,-0.33) & (-0.17,-0.70) & ( 0.23,-1.24) & ( 0.49,-0.26) & ( 0.63,-0.18) & ( 0.54,-0.09) & ( 0.59, 0.17) & ( 0.50, 0.18) & ( 1.67, 2.25) \\ 
  23 & WTF & (-1.70, 1.14) & (-0.42, 0.05) & (-0.58,-1.01) & (-0.32,-0.63) & (-0.26,-0.53) & ( 0.22,-0.65) & ( 0.46,-0.39) & ( 0.61,-0.19) & ( 0.61, 0.54) & ( 0.51, 0.61) & ( 0.61, 0.46) & ( 0.74, 0.79) \\ 
   \hline
\end{tabular}
}
    \caption{Coordinates obtained using Correspondence Analysis for Subverses with size at least $1500$.}
    \label{tab:CA_voat}
\end{table}

\begin{table}[!ht]
\vskip-25pt
\hspace{-25mm}
\scalebox{0.6}{
\begin{tabular}{rlllllllllllll}
  \hline
 & Subreddit & 01-2019 & 02-2019 & 03-2019 & 04-2019 & 05-2019 & 06-2019 & 07-2019 & 08-2019 & 09-2019 & 10-2019 & 11-2019 & 12-2019 \\ 
  \hline
1 & memes & ( 2.11,-1.11) & ( 1.66,-0.46) & ( 1.23, 0.11) & ( 0.84, 0.49) & ( 0.52, 0.65) & ( 0.20, 0.91) & ( 0.01, 0.83) & (-0.18, 0.55) & (-0.34, 0.17) & (-0.49,-0.21) & (-0.62,-0.61) & (-0.73,-1.03) \\ 
  2 & gaming & ( 1.58,-0.69) & ( 1.03,-0.10) & ( 0.78, 0.23) & ( 0.50, 0.45) & ( 0.25, 0.52) & ( 0.02, 0.93) & (-0.15, 0.86) & (-0.32, 0.58) & (-0.43, 0.16) & (-0.57,-0.23) & (-0.68,-0.70) & (-0.82,-1.23) \\ 
  3 & aww & ( 1.62,-0.73) & ( 1.17,-0.20) & ( 0.88, 0.22) & ( 0.58, 0.49) & ( 0.31, 0.58) & ( 0.04, 0.98) & (-0.13, 0.93) & (-0.30, 0.63) & (-0.45, 0.19) & (-0.60,-0.26) & (-0.73,-0.75) & (-0.84,-1.24) \\ 
  4 & dankmemes & ( 1.84, 0.92) & ( 1.42, 0.38) & ( 1.10,-0.02) & ( 0.72,-0.36) & ( 0.44,-0.54) & ( 0.14,-0.81) & (-0.03,-0.76) & (-0.21,-0.54) & (-0.37,-0.14) & (-0.52, 0.26) & (-0.65, 0.65) & (-0.76, 1.06) \\ 
  5 & Showerthoughts & ( 1.65,-0.69) & ( 1.15,-0.16) & ( 0.87, 0.24) & ( 0.55, 0.52) & ( 0.28, 0.63) & ( 0.01, 0.98) & (-0.16, 0.89) & (-0.32, 0.56) & (-0.45, 0.16) & (-0.57,-0.25) & (-0.70,-0.73) & (-0.81,-1.27) \\ 
  6 & mildlyinteresting & ( 1.74, 0.69) & ( 1.22, 0.27) & ( 0.87,-0.05) & ( 0.55,-0.30) & ( 0.30,-0.45) & ( 0.04,-1.25) & (-0.16,-0.73) & (-0.34,-0.37) & (-0.46, 0.05) & (-0.59, 0.42) & (-0.69, 0.75) & (-0.82, 1.21) \\ 
  7 & worldnews & ( 1.53,-0.80) & ( 1.03,-0.15) & ( 0.84, 0.09) & ( 0.65, 0.29) & ( 0.44, 0.37) & ( 0.15, 0.83) & (-0.02, 0.87) & (-0.21, 0.60) & (-0.34, 0.23) & (-0.54,-0.15) & (-0.60,-0.49) & (-0.77,-1.06) \\ 
  8 & todayilearned & ( 1.58,-0.59) & ( 0.88, 0.07) & ( 0.67, 0.32) & ( 0.47, 0.51) & ( 0.25, 0.49) & (-0.03, 0.90) & (-0.19, 0.85) & (-0.32, 0.49) & (-0.41, 0.10) & (-0.51,-0.31) & (-0.61,-0.73) & (-0.68,-1.15) \\ 
  9 & PewdiepieSubmissions & ( 1.84,-0.78) & ( 1.40, 0.19) & ( 1.01, 0.84) & ( 0.65, 1.20) & ( 0.31, 0.82) & (-0.03, 0.82) & (-0.33, 0.73) & (-0.48, 0.27) & (-0.51,-0.26) & (-0.53,-0.57) & (-0.54,-0.79) & (-0.57,-1.02) \\ 
  10 & gifs & (-1.52,-0.56) & (-0.90, 0.19) & (-0.63, 0.63) & (-0.37, 0.91) & (-0.11, 0.88) & ( 0.15, 1.01) & ( 0.29, 0.77) & ( 0.46, 0.55) & ( 0.52, 0.02) & ( 0.64,-0.46) & ( 0.65,-0.82) & ( 0.73,-1.31) \\ 
  11 & videos & ( 1.64,-0.47) & ( 0.84, 0.14) & ( 0.64, 0.41) & ( 0.34, 0.46) & ( 0.17, 0.46) & (-0.13, 1.17) & (-0.28, 0.69) & (-0.39, 0.34) & (-0.46,-0.12) & (-0.49,-0.43) & (-0.56,-0.81) & (-0.61,-1.15) \\ 
  12 & RoastMe & ( 1.87,-0.68) & ( 1.36,-0.09) & ( 0.72, 0.42) & ( 0.48, 0.82) & ( 0.22, 0.85) & (-0.03, 1.15) & (-0.20, 0.87) & (-0.40, 0.53) & (-0.49, 0.08) & (-0.57,-0.32) & (-0.69,-0.83) & (-0.78,-1.32) \\ 
  13 & relationship\_advice & ( 2.07,-0.95) & ( 1.41,-0.30) & ( 1.05, 0.20) & ( 0.73, 0.49) & ( 0.42, 0.61) & ( 0.10, 0.98) & (-0.09, 0.95) & (-0.28, 0.61) & (-0.44, 0.16) & (-0.59,-0.36) & (-0.72,-0.83) & (-0.83,-1.31) \\ 
  14 & movies & ( 1.62,-0.76) & ( 0.84,-0.09) & ( 0.64, 0.09) & ( 0.44, 0.23) & ( 0.34, 0.41) & ( 0.11, 0.73) & (-0.04, 0.98) & (-0.19, 0.51) & (-0.29, 0.24) & (-0.58,-0.07) & (-0.57,-0.35) & (-0.84,-1.26) \\ 
  15 & news & ( 1.88, 1.25) & ( 1.73, 1.09) & ( 1.39, 0.72) & ( 1.16, 0.41) & ( 0.87, 0.20) & ( 0.39,-0.43) & ( 0.20,-0.62) & (-0.09,-0.90) & (-0.23,-0.28) & (-0.51, 0.09) & (-0.59, 0.44) & (-0.87, 1.08) \\ 
  16 & trashy & ( 1.71, 0.83) & ( 1.24, 0.31) & ( 0.98,-0.06) & ( 0.69,-0.34) & ( 0.41,-0.50) & ( 0.13,-0.97) & (-0.08,-0.92) & (-0.28,-0.56) & (-0.45,-0.13) & (-0.62, 0.34) & (-0.73, 0.74) & (-0.84, 1.19) \\ 
  17 & interestingasfuck & ( 1.74,-0.92) & ( 1.16, 1.05) & ( 0.76, 1.98) & ( 0.37, 1.66) & ( 0.11, 1.16) & (-0.13, 0.38) & (-0.26, 0.22) & (-0.39, 0.01) & (-0.47,-0.17) & (-0.57,-0.36) & (-0.56,-0.43) & (-0.63,-0.61) \\ 
  18 & wholesomememes & ( 1.93, 0.80) & ( 1.59, 0.41) & ( 1.18,-0.07) & ( 0.71,-0.46) & ( 0.39,-0.71) & ( 0.04,-1.01) & (-0.16,-0.85) & (-0.33,-0.54) & (-0.45,-0.11) & (-0.59, 0.37) & (-0.74, 0.91) & (-0.88, 1.55) \\ 
  19 & PublicFreakout & ( 2.08, 1.24) & ( 1.68, 0.66) & ( 1.29, 0.25) & ( 1.05,-0.11) & ( 0.79,-0.25) & ( 0.40,-0.76) & ( 0.17,-0.86) & (-0.08,-0.75) & (-0.30,-0.32) & (-0.50, 0.16) & (-0.66, 0.64) & (-0.77, 1.07) \\ 
  20 & Minecraft & ( 2.42, 2.57) & ( 2.05, 1.97) & ( 1.72, 1.41) & ( 1.39, 0.83) & ( 1.12, 0.30) & ( 0.75,-0.52) & ( 0.45,-0.77) & ( 0.08,-0.60) & (-0.28,-0.19) & (-0.58, 0.21) & (-0.82, 0.57) & (-1.03, 0.93) \\ 
  21 & WTF & ( 1.56,-0.81) & ( 1.03,-0.25) & ( 0.78, 0.06) & ( 0.63, 0.28) & ( 0.36, 0.43) & ( 0.11, 0.82) & (-0.07, 0.85) & (-0.25, 0.65) & (-0.39, 0.24) & (-0.58,-0.17) & (-0.72,-0.65) & (-0.88,-1.23) \\ 
  22 & oddlysatisfying & ( 1.80,-0.76) & ( 1.24,-0.21) & ( 0.94, 0.20) & ( 0.58, 0.56) & ( 0.31, 0.72) & ( 0.06, 1.00) & (-0.11, 1.02) & (-0.28, 0.64) & (-0.43, 0.25) & (-0.57,-0.20) & (-0.71,-0.70) & (-0.84,-1.35) \\ 
  23 & Wellthatsucks & ( 2.00,-0.83) & ( 1.44,-0.28) & ( 1.11, 0.22) & ( 0.79, 0.50) & ( 0.44, 0.50) & ( 0.14, 0.76) & (-0.05, 1.14) & (-0.29, 0.62) & (-0.44, 0.14) & (-0.58,-0.33) & (-0.61,-0.71) & (-0.73,-1.31) \\ 
  24 & tifu & ( 1.92, 0.89) & ( 1.34, 0.38) & ( 0.94,-0.03) & ( 0.70,-0.43) & ( 0.40,-0.58) & ( 0.14,-1.17) & (-0.06,-0.84) & (-0.25,-0.50) & (-0.40,-0.03) & (-0.65, 0.51) & (-0.70, 0.79) & (-0.79, 1.08) \\ 
  25 & AskMen & ( 1.70,-0.81) & ( 1.30,-0.26) & ( 1.19, 0.16) & ( 0.57, 0.34) & ( 0.36, 0.49) & ( 0.10, 0.85) & (-0.07, 0.83) & (-0.26, 0.75) & (-0.45, 0.45) & (-0.56, 0.00) & (-0.71,-0.57) & (-0.89,-1.46) \\ 
  26 & mildlyinfuriating & ( 1.88,-0.86) & ( 1.47,-0.34) & ( 1.10, 0.15) & ( 0.69, 0.39) & ( 0.45, 0.58) & ( 0.14, 0.95) & (-0.08, 0.92) & (-0.26, 0.68) & (-0.43, 0.23) & (-0.58,-0.22) & (-0.68,-0.64) & (-0.84,-1.39) \\ 
  27 & ChoosingBeggars & ( 1.57,-0.60) & ( 1.35,-0.32) & ( 0.73, 0.22) & ( 0.41, 0.43) & ( 0.20, 0.61) & (-0.05, 0.97) & (-0.20, 0.97) & (-0.34, 0.68) & (-0.44, 0.27) & (-0.57,-0.14) & (-0.87,-1.17) & (-0.81,-0.97) \\ 
  28 & BlackPeopleTwitter & ( 1.64,-0.58) & ( 1.04,-0.08) & ( 0.76, 0.16) & ( 0.47, 0.41) & ( 0.23, 0.47) & (-0.01, 0.95) & (-0.20, 0.89) & (-0.40, 0.62) & (-0.51, 0.09) & (-0.61,-0.39) & (-0.71,-0.97) & (-0.74,-1.32) \\ 
  29 & insanepeoplefacebook & ( 1.80,-1.02) & ( 1.41,-0.45) & ( 1.05,-0.02) & ( 0.81, 0.34) & ( 0.53, 0.46) & ( 0.23, 1.06) & ( 0.02, 0.86) & (-0.21, 0.66) & (-0.36, 0.24) & (-0.57,-0.20) & (-0.74,-0.72) & (-0.83,-1.03) \\ 
  30 & me\_irl & ( 1.42, 0.77) & ( 1.10, 0.43) & ( 0.99, 0.24) & ( 0.67,-0.05) & ( 0.41,-0.28) & ( 0.18,-0.74) & (-0.04,-0.83) & (-0.30,-0.88) & (-0.48,-0.33) & (-0.71, 0.15) & (-0.93, 0.76) & (-1.10, 1.41) \\ 
  31 & OldSchoolCool & ( 1.78,-0.65) & ( 1.05, 0.01) & ( 0.69, 0.32) & ( 0.49, 0.59) & ( 0.29, 0.77) & (-0.02, 1.17) & (-0.20, 0.93) & (-0.36, 0.49) & (-0.43, 0.08) & (-0.59,-0.46) & (-0.68,-0.85) & (-0.73,-1.19) \\ 
  32 & nextfuckinglevel & ( 2.53, 9.24) & ( 3.15,10.95) & ( 2.08, 5.34) & ( 2.14, 5.74) & ( 1.58, 1.28) & ( 1.44,-0.15) & ( 1.13,-0.32) & ( 0.66,-0.37) & ( 0.25,-0.33) & (-0.17,-0.14) & (-0.59, 0.10) & (-1.08, 0.35) \\ 
  33 & Damnthatsinteresting & ( 1.88,-0.73) & ( 1.07,-0.08) & ( 0.80, 0.26) & ( 0.51, 0.53) & ( 0.29, 0.66) & ( 0.04, 1.07) & (-0.16, 0.96) & (-0.32, 0.50) & (-0.44, 0.09) & (-0.57,-0.33) & (-0.74,-0.92) & (-0.79,-1.29) \\ 
  34 & NoStupidQuestions & ( 1.88,-0.71) & ( 1.08,-0.03) & ( 0.78, 0.30) & ( 0.52, 0.46) & ( 0.33, 0.80) & ( 0.07, 0.89) & (-0.11, 0.92) & (-0.32, 0.75) & (-0.47, 0.17) & (-0.58,-0.36) & (-0.66,-0.81) & (-0.75,-1.32) \\ 
  35 & Whatcouldgowrong & ( 1.93, 1.17) & ( 1.57, 0.63) & ( 1.04, 0.02) & ( 0.85,-0.32) & ( 0.57,-0.55) & ( 0.28,-1.05) & ( 0.06,-0.92) & (-0.12,-0.57) & (-0.36,-0.18) & (-0.57, 0.22) & (-0.69, 0.58) & (-0.88, 1.09) \\ 
  36 & MurderedByWords & ( 1.88, 0.68) & ( 1.29, 0.18) & ( 0.95,-0.18) & ( 0.64,-0.50) & ( 0.28,-0.51) & ( 0.01,-0.97) & (-0.17,-0.90) & (-0.35,-0.54) & (-0.45,-0.06) & (-0.57, 0.43) & (-0.67, 0.86) & (-0.75, 1.36) \\ 
  37 & nba & (-0.05, 1.24) & ( 0.01, 0.90) & ( 0.06, 0.79) & ( 0.05, 0.67) & (-0.09, 0.52) & (-1.15,-0.47) & (-0.38,-0.26) & ( 0.17,-0.16) & ( 0.41,-0.19) & ( 0.78,-0.62) & ( 0.62,-0.33) & ( 0.65,-0.29) \\ 
  38 & LifeProTips & ( 1.83,-0.92) & ( 1.55,-0.42) & ( 0.94, 0.13) & ( 0.80, 0.43) & ( 0.48, 0.55) & ( 0.23, 0.89) & ( 0.02, 0.93) & (-0.17, 0.82) & (-0.35, 0.50) & (-0.50, 0.08) & (-0.67,-0.50) & (-0.91,-1.34) \\ 
  39 & cursedcomments & ( 7.80, 5.11) & ( 7.07, 4.33) & ( 4.56, 1.86) & ( 2.89, 0.27) & ( 1.27,-0.67) & ( 0.55,-1.02) & ( 0.34,-0.82) & ( 0.12,-0.48) & (-0.10,-0.10) & (-0.33, 0.30) & (-0.49, 0.62) & (-0.68, 1.03) \\ 
  40 & gonewild & ( 1.77,-0.86) & ( 1.26,-0.30) & ( 0.94, 0.09) & ( 0.66, 0.41) & ( 0.40, 0.57) & ( 0.10, 0.99) & (-0.09, 0.88) & (-0.25, 0.59) & (-0.43, 0.19) & (-0.58,-0.23) & (-0.72,-0.71) & (-0.85,-1.19) \\ 
  41 & NatureIsFuckingLit & ( 2.01, 1.20) & ( 1.52, 0.59) & ( 1.08, 0.06) & ( 0.74,-0.27) & ( 0.64,-0.85) & ( 0.28,-1.10) & ( 0.05,-0.84) & (-0.18,-0.60) & (-0.39,-0.18) & (-0.57, 0.19) & (-0.81, 0.74) & (-0.92, 1.16) \\ 
  42 & Unexpected & ( 1.97,-0.88) & ( 1.31,-0.16) & ( 1.05, 0.37) & ( 0.59, 0.48) & ( 0.34, 0.57) & ( 0.12, 1.00) & (-0.05, 0.95) & (-0.23, 0.72) & (-0.37, 0.22) & (-0.53,-0.23) & (-0.71,-0.79) & (-0.83,-1.29) \\ 
  43 & WhitePeopleTwitter & ( 1.73,-0.66) & ( 1.09, 0.17) & ( 0.70, 0.50) & ( 0.46, 0.73) & ( 0.22, 0.72) & (-0.04, 1.03) & (-0.19, 0.89) & (-0.32, 0.58) & (-0.45, 0.21) & (-0.54,-0.22) & (-0.67,-0.75) & (-0.75,-1.27) \\ 
  44 & IdiotsInCars & ( 1.95, 1.53) & ( 1.43, 0.70) & ( 1.48, 0.49) & ( 1.02,-0.02) & ( 0.78,-0.30) & ( 0.52,-0.81) & ( 0.32,-0.99) & ( 0.07,-0.66) & (-0.21,-0.47) & (-0.44,-0.07) & (-0.71, 0.45) & (-0.90, 0.95) \\ 
  45 & AskOuija & ( 1.73,-0.67) & ( 1.27,-0.06) & ( 0.82, 0.55) & ( 0.46, 0.75) & ( 0.15, 0.82) & (-0.08, 1.06) & (-0.21, 0.87) & (-0.39, 0.59) & (-0.50, 0.11) & (-0.63,-0.35) & (-0.74,-0.83) & (-0.83,-1.34) \\ 
  46 & Tinder & ( 1.84,-0.81) & ( 1.27,-0.31) & ( 0.93, 0.05) & ( 0.64, 0.35) & ( 0.40, 0.53) & ( 0.09, 1.00) & (-0.10, 0.94) & (-0.26, 0.66) & (-0.43, 0.24) & (-0.61,-0.22) & (-0.73,-0.70) & (-0.94,-1.44) \\ 
  47 & iamatotalpieceofshit & ( 2.00,-1.00) & ( 1.52,-0.34) & ( 1.35, 0.30) & ( 0.75, 0.49) & ( 0.46, 0.66) & ( 0.12, 1.14) & (-0.08, 0.85) & (-0.26, 0.52) & (-0.39, 0.13) & (-0.55,-0.30) & (-0.67,-0.69) & (-0.79,-1.19) \\ 
  48 & facepalm & ( 1.98,-0.94) & ( 1.43,-0.32) & ( 1.01, 0.24) & ( 0.64, 0.46) & ( 0.44, 0.65) & ( 0.14, 1.06) & (-0.06, 0.96) & (-0.22, 0.60) & (-0.36, 0.24) & (-0.54,-0.18) & (-0.66,-0.60) & (-0.86,-1.25) \\ 
  49 & blursedimages & ( 4.55,10.96) & ( 4.55,10.48) & ( 3.88, 8.02) & ( 3.45, 5.28) & ( 2.40, 0.80) & ( 1.50,-0.50) & ( 1.10,-0.58) & ( 0.66,-0.46) & ( 0.19,-0.27) & (-0.21,-0.04) & (-0.61, 0.20) & (-0.94, 0.44) \\ 
  50 & HistoryMemes & ( 1.89,-1.10) & ( 1.37,-0.43) & ( 1.10,-0.01) & ( 0.87, 0.44) & ( 0.51, 0.54) & ( 0.21, 0.97) & (-0.01, 0.83) & (-0.19, 0.56) & (-0.37, 0.20) & (-0.54,-0.15) & (-0.71,-0.61) & (-0.85,-1.11) \\ 
  51 & nfl & ( 0.98,-0.80) & ( 1.33, 0.26) & ( 0.43, 0.88) & ( 0.31, 0.98) & ( 0.24, 1.03) & ( 0.09, 0.98) & ( 0.01, 0.87) & (-0.13, 0.62) & (-0.35,-0.06) & (-0.44,-0.15) & (-0.53,-0.32) & (-0.57,-0.42) \\ 
  52 & WatchPeopleDieInside & ( 2.41,-1.32) & ( 1.73,-0.28) & ( 1.35, 0.12) & ( 0.96, 0.51) & ( 0.68, 0.70) & ( 0.36, 1.09) & ( 0.17, 0.92) & (-0.04, 0.80) & (-0.23, 0.38) & (-0.44,-0.06) & (-0.62,-0.60) & (-0.76,-1.10) \\ 
  53 & science & ( 1.92,-1.09) & ( 1.60,-0.65) & ( 0.99, 0.09) & ( 1.25, 2.13) & ( 0.48, 0.47) & ( 0.12, 0.79) & (-0.04, 0.94) & (-0.20, 0.45) & (-0.40, 0.16) & (-0.56,-0.29) & (-0.65,-0.58) & (-0.80,-1.10) \\ 
  54 & CrappyDesign & ( 1.90, 0.60) & ( 1.34, 0.15) & ( 0.93,-0.14) & ( 0.53,-0.38) & ( 0.28,-0.53) & ( 0.02,-0.99) & (-0.20,-0.95) & (-0.40,-0.65) & (-0.50, 0.02) & (-0.61, 0.55) & (-0.67, 0.97) & (-0.74, 1.41) \\ 
  55 & NintendoSwitch & (-1.69,-0.23) & (-0.22, 0.36) & (-0.03, 0.50) & ( 0.09, 0.53) & ( 0.16, 0.45) & ( 0.21, 0.86) & ( 0.34, 0.86) & ( 0.37, 0.38) & ( 0.42, 0.13) & ( 0.51,-0.30) & ( 0.54,-0.82) & ( 0.59,-1.47) \\ 
  56 & starterpacks & ( 1.71,-0.83) & ( 1.28,-0.08) & ( 0.83, 0.29) & ( 0.56, 0.45) & ( 0.33, 0.48) & ( 0.15, 1.14) & (-0.07, 0.92) & (-0.24, 0.69) & (-0.37, 0.24) & (-0.67,-0.32) & (-0.71,-0.71) & (-0.77,-1.12) \\ 
  57 & Jokes & ( 1.81,-0.60) & ( 1.11, 0.15) & ( 0.75, 0.54) & ( 0.52, 1.03) & ( 0.18, 0.80) & (-0.12, 1.10) & (-0.28, 0.88) & (-0.42, 0.44) & (-0.50,-0.06) & (-0.59,-0.51) & (-0.65,-0.87) & (-0.70,-1.34) \\ 
  58 & nottheonion & ( 1.88,-0.70) & ( 1.15, 0.11) & ( 0.86, 0.43) & ( 0.55, 0.70) & ( 0.29, 0.47) & ( 0.01, 1.04) & (-0.19, 0.96) & (-0.29, 0.53) & (-0.39, 0.14) & (-0.53,-0.30) & (-0.60,-0.75) & (-0.69,-1.21) \\ 
  59 & HumansBeingBros & ( 1.86,-0.93) & ( 1.40,-0.21) & ( 1.05, 0.46) & ( 0.76, 0.92) & ( 0.33, 0.71) & ( 0.06, 1.00) & (-0.10, 0.95) & (-0.29, 0.64) & (-0.44, 0.20) & (-0.58,-0.20) & (-0.72,-0.74) & (-0.86,-1.33) \\ 
  60 & BeAmazed & ( 1.69,-0.63) & ( 1.17,-0.14) & ( 0.89, 0.37) & ( 0.45, 0.53) & ( 0.19, 0.59) & (-0.04, 0.98) & (-0.22, 0.95) & (-0.35, 0.79) & (-0.46, 0.28) & (-0.64,-0.23) & (-0.75,-0.84) & (-0.87,-1.52) \\ 
  61 & therewasanattempt & ( 2.05,-1.31) & ( 1.59,-0.55) & ( 1.20, 0.20) & ( 1.20, 1.24) & ( 0.66, 1.84) & ( 0.22, 1.07) & ( 0.01, 0.78) & (-0.17, 0.50) & (-0.32, 0.13) & (-0.52,-0.24) & (-0.69,-0.60) & (-0.85,-0.94) \\ 
  62 & technology & ( 1.76,-0.64) & ( 0.85, 0.11) & ( 0.70, 0.45) & ( 0.37, 0.42) & ( 0.19, 0.50) & ( 0.02, 0.90) & (-0.15, 0.99) & (-0.28, 0.57) & (-0.41, 0.17) & (-0.56,-0.29) & (-0.63,-0.72) & (-0.73,-1.29) \\ 
  63 & Cringetopia & ( 2.34, 1.45) & ( 1.94, 0.90) & ( 1.46, 0.31) & ( 1.10,-0.13) & ( 0.80,-0.36) & ( 0.50,-0.96) & ( 0.26,-0.86) & ( 0.01,-0.72) & (-0.23,-0.41) & (-0.48, 0.04) & (-0.60, 0.48) & (-0.78, 1.03) \\ 
  64 & DestinyTheGame & ( 0.59, 1.09) & ( 0.55, 0.88) & ( 0.48, 0.73) & ( 0.46, 0.65) & ( 0.43, 0.58) & ( 0.35, 0.50) & ( 0.29, 0.34) & ( 0.17, 0.18) & (-1.52, 0.15) & ( 0.23,-1.13) & ( 0.33,-0.60) & ( 0.38,-0.52) \\ 
  65 & assholedesign & ( 1.82,-0.67) & ( 1.19,-0.13) & ( 0.80, 0.29) & ( 0.56, 0.84) & ( 0.22, 0.56) & (-0.02, 1.11) & (-0.21, 0.95) & (-0.34, 0.52) & (-0.49, 0.09) & (-0.57,-0.36) & (-0.74,-1.03) & (-0.75,-1.22) \\ 
  66 & Music & ( 1.73,-0.68) & ( 1.10,-0.06) & ( 0.84, 0.54) & ( 0.29, 0.33) & ( 0.13, 0.46) & (-0.06, 1.07) & (-0.17, 0.95) & (-0.34, 0.68) & (-0.46, 0.15) & (-0.69,-0.35) & (-0.71,-0.64) & (-0.92,-1.65) \\ 
  67 & IAmA & ( 2.09,-0.75) & ( 1.54,-0.38) & ( 0.68, 0.33) & ( 0.61, 1.27) & ( 0.49, 1.18) & ( 0.07, 1.84) & (-0.09, 0.90) & (-0.26, 0.58) & (-0.50,-0.16) & (-0.77,-1.00) & (-0.56,-0.51) & (-0.62,-0.54) \\ 
  68 & soccer & ( 0.81, 1.08) & ( 0.55, 0.73) & ( 0.49, 0.61) & ( 0.39, 0.46) & ( 0.35, 0.36) & ( 0.57,-0.69) & ( 0.48,-1.01) & (-0.08,-0.20) & (-0.32,-0.04) & (-0.54, 0.03) & (-0.79, 0.10) & (-0.93, 0.16) \\ 
  69 & KidsAreFuckingStupid & ( 2.26, 1.00) & ( 1.91, 0.57) & ( 1.07,-0.11) & ( 0.76,-0.39) & ( 0.47,-0.56) & ( 0.16,-1.16) & (-0.02,-0.88) & (-0.19,-0.57) & (-0.34,-0.17) & (-0.48, 0.20) & (-0.68, 0.78) & (-0.82, 1.28) \\ 
  70 & food & ( 1.95,-0.69) & ( 1.38,-0.09) & ( 0.86, 0.40) & ( 0.48, 0.67) & ( 0.24, 0.79) & (-0.04, 1.17) & (-0.20, 0.93) & (-0.33, 0.54) & (-0.46, 0.13) & (-0.59,-0.37) & (-0.64,-0.78) & (-0.79,-1.42) \\ 
  71 & PoliticalHumor & ( 1.73,-0.51) & ( 1.03, 0.08) & ( 0.55, 0.22) & ( 0.38, 0.41) & ( 0.17, 0.42) & (-0.05, 0.92) & (-0.23, 0.99) & (-0.36, 0.57) & (-0.44, 0.02) & (-0.58,-0.47) & (-0.61,-0.82) & (-0.70,-1.26) \\ 
  72 & insaneparents & ( 3.42, 5.61) & ( 3.45, 5.21) & ( 2.79, 4.06) & ( 2.45, 2.98) & ( 1.91, 1.09) & ( 1.43, 0.05) & ( 1.14,-0.25) & ( 0.68,-0.56) & ( 0.06,-0.68) & (-0.36,-0.10) & (-0.70, 0.38) & (-1.03, 0.88) \\ 
  73 & DnD & ( 0.87, 1.61) & ( 0.81, 1.08) & ( 0.75, 0.70) & ( 0.75, 0.55) & ( 0.66, 0.22) & ( 0.57, 0.02) & ( 0.56,-0.20) & ( 0.46,-0.40) & ( 0.30,-0.47) & (-1.32, 0.26) & ( 0.15,-0.97) & ( 0.33,-1.29) \\ 
  74 & marvelstudios & ( 1.74, 0.63) & ( 0.83, 0.08) & ( 0.66,-0.07) & ( 0.56,-0.55) & ( 0.23,-1.03) & (-0.15,-0.70) & (-0.32,-0.63) & (-0.52,-0.12) & (-0.50, 0.38) & (-0.57, 0.75) & (-0.62, 1.14) & (-0.59, 1.21) \\ 
  75 & PrequelMemes & ( 1.54,-0.87) & ( 1.05,-0.22) & ( 0.77, 0.14) & ( 0.52, 0.36) & ( 0.37, 0.57) & ( 0.17, 0.81) & (-0.01, 0.89) & (-0.18, 0.68) & (-0.31, 0.47) & (-0.48, 0.21) & (-0.66,-0.25) & (-0.98,-1.15) \\ 
  76 & MadeMeSmile & ( 1.96,-0.82) & ( 1.46,-0.09) & ( 1.04, 0.60) & ( 0.59, 0.75) & ( 0.37, 0.82) & ( 0.05, 1.02) & (-0.12, 0.90) & (-0.26, 0.71) & (-0.39, 0.35) & (-0.52,-0.08) & (-0.69,-0.70) & (-0.84,-1.47) \\ 
  77 & wallstreetbets & ( 0.44, 0.73) & ( 0.43, 0.50) & ( 0.37, 0.40) & (-1.50,-0.09) & ( 0.39, 0.34) & ( 0.47, 1.68) & ( 0.53, 0.32) & ( 0.55,-0.09) & ( 0.61,-0.34) & ( 0.63,-0.62) & ( 0.66,-1.12) & ( 0.67,-1.03) \\ 
  78 & relationships & ( 1.68,-0.64) & ( 1.04,-0.11) & ( 0.71, 0.19) & ( 0.50, 0.39) & ( 0.27, 0.49) & ( 0.01, 1.06) & (-0.18, 0.94) & (-0.36, 0.62) & (-0.48, 0.17) & (-0.59,-0.30) & (-0.70,-0.76) & (-0.82,-1.28) \\ 
  79 & JusticeServed & ( 1.85, 1.28) & ( 1.40, 0.70) & ( 1.00, 0.22) & ( 0.90,-0.08) & ( 0.63,-0.27) & ( 0.41,-0.94) & ( 0.21,-1.08) & (-0.03,-0.58) & (-0.32,-0.29) & (-0.88, 0.26) & (-0.86, 0.96) & (-0.76, 0.84) \\ 
  80 & AdviceAnimals & (-1.65,-0.48) & (-0.82, 0.10) & (-0.56, 0.35) & (-0.33, 0.42) & (-0.14, 0.50) & ( 0.09, 0.93) & ( 0.23, 0.92) & ( 0.35, 0.56) & ( 0.46, 0.22) & ( 0.55,-0.12) & ( 0.60,-0.51) & ( 0.77,-1.62) \\ 
  81 & madlads & ( 2.35,-1.12) & ( 1.75,-0.38) & ( 1.47, 0.37) & ( 0.93, 0.67) & ( 0.55, 0.77) & ( 0.24, 1.19) & ( 0.02, 0.97) & (-0.13, 0.61) & (-0.31, 0.24) & (-0.51,-0.18) & (-0.62,-0.61) & (-0.79,-1.24) \\ 
  82 & TooAfraidToAsk & (-1.83,-0.47) & (-0.98, 0.31) & (-0.52, 0.70) & (-0.24, 0.85) & (-0.06, 0.82) & ( 0.18, 1.00) & ( 0.31, 0.83) & ( 0.50, 0.55) & ( 0.54,-0.11) & ( 0.58,-0.56) & ( 0.63,-0.95) & ( 0.69,-1.50) \\ 
  83 & explainlikeimfive & ( 1.97,-0.44) & ( 1.09,-0.04) & ( 0.65, 0.20) & ( 0.42, 0.35) & ( 0.20, 0.58) & (-0.10, 1.18) & (-0.25, 0.89) & (-0.39, 0.66) & (-0.46, 0.00) & (-0.54,-0.55) & (-0.56,-0.84) & (-0.64,-1.49) \\ 
  84 & millionairemakers & (  2.20, -0.07) & (  3.13, -0.12) & (  6.35,274.28) & (  2.59, -0.11) & (  2.64, -0.09) & (  2.15, -0.06) & (  2.13, -0.06) & (  2.12, -0.06) & (  2.11,  0.01) & (  2.04, -0.05) & (  1.99, -0.05) & ( -0.39,  0.01) \\ 
  85 & xboxone & ( 0.70, 1.76) & ( 0.63, 0.85) & ( 0.54, 0.44) & ( 0.52, 0.23) & ( 0.52, 0.06) & ( 0.49,-0.24) & ( 0.43,-0.35) & ( 0.40,-0.51) & ( 0.33,-0.63) & ( 0.23,-0.69) & (-0.03,-0.64) & (-1.55, 0.26) \\ 
  86 & books & (-1.72,-0.46) & (-0.61, 0.45) & (-0.40, 1.19) & (-0.12, 0.82) & ( 0.02, 0.79) & ( 0.18, 0.96) & ( 0.29, 0.96) & ( 0.38, 0.67) & ( 0.47, 0.17) & ( 0.78,-1.51) & ( 0.64,-0.61) & ( 0.59,-0.26) \\ 
  87 & blackmagicfuckery & ( 1.97, 0.97) & ( 1.48, 0.45) & ( 1.21, 0.02) & ( 0.78,-0.42) & ( 0.44,-0.57) & ( 0.11,-1.09) & (-0.11,-0.92) & (-0.25,-0.55) & (-0.34,-0.15) & (-0.59, 0.28) & (-0.81, 0.83) & (-0.94, 1.33) \\ 
  88 & apolloapp & ( 2.56, 4.49) & ( 2.45, 2.93) & ( 2.38, 1.48) & ( 2.46, 1.19) & ( 2.39, 0.59) & ( 2.39, 0.26) & ( 2.29,-0.11) & ( 2.23,-0.28) & (-0.37, 0.01) & ( 2.17,-1.91) & ( 2.29,-2.12) & ( 2.34,-2.36) \\ 
  89 & natureismetal & ( 2.04, 1.03) & ( 1.63, 0.57) & ( 1.13, 0.15) & ( 0.73,-0.16) & ( 0.52,-0.46) & ( 0.21,-1.02) & (-0.01,-0.90) & (-0.29,-0.62) & (-0.46,-0.05) & (-0.57, 0.38) & (-0.84, 1.06) & (-0.84, 1.15) \\ 
  90 & cursedimages & ( 2.03,-1.14) & ( 1.66,-0.63) & ( 1.43,-0.06) & ( 0.96, 0.36) & ( 0.72, 0.89) & ( 0.34, 1.14) & ( 0.14, 1.11) & (-0.10, 0.89) & (-0.34, 0.42) & (-0.53,-0.07) & (-0.70,-0.56) & (-0.88,-1.13) \\ 
  91 & Rainbow6 & ( 1.61, 0.89) & ( 1.24, 0.47) & ( 0.93, 0.11) & ( 0.66,-0.16) & ( 0.41,-0.36) & ( 0.22,-0.85) & ( 0.01,-0.87) & (-0.22,-0.62) & (-0.43,-0.23) & (-0.62, 0.21) & (-0.75, 0.62) & (-0.96, 1.21) \\ 
  92 & instantkarma & ( 2.98,-0.97) & ( 1.45,-0.02) & ( 1.16, 0.16) & ( 0.88, 0.60) & ( 0.51, 0.67) & ( 0.23, 1.15) & ( 0.05, 1.07) & (-0.16, 0.65) & (-0.30, 0.13) & (-0.40,-0.30) & (-0.52,-0.85) & (-0.60,-1.22) \\
  93 & trees & ( 1.61,-0.68) & ( 1.15,-0.17) & ( 0.83, 0.21) & ( 0.55, 0.53) & ( 0.31, 0.71) & ( 0.01, 0.98) & (-0.18, 0.89) & (-0.33, 0.57) & (-0.46, 0.21) & (-0.60,-0.25) & (-0.71,-0.77) & (-0.83,-1.23) \\ 
  94 & classicwow & ( 2.02, 1.29) & ( 2.00, 1.28) & ( 1.90, 1.16) & ( 1.71, 0.98) & ( 1.50, 0.77) & ( 1.12, 0.17) & ( 0.93, 0.10) & ( 0.37,-0.90) & (-0.42,-0.54) & (-0.57, 0.34) & (-0.56, 0.74) & (-0.55, 0.93) \\ 
  95 & PS4 & ( 1.61,-1.02) & ( 0.81,-0.23) & ( 0.63,-0.01) & ( 0.47, 0.13) & ( 0.31, 0.27) & ( 0.18, 0.79) & ( 0.03, 0.74) & (-0.14, 0.53) & (-0.28, 0.46) & (-0.42, 0.23) & (-0.58,-0.14) & (-1.18,-1.36) \\ 
  96 & buildapc & ( 1.78,-1.07) & ( 1.30,-0.52) & ( 1.07,-0.15) & ( 0.71, 0.25) & ( 0.46, 0.50) & ( 0.26, 0.96) & ( 0.08, 1.04) & (-0.14, 0.82) & (-0.34, 0.41) & (-0.57,-0.05) & (-0.84,-0.66) & (-1.02,-1.15) \\ 
  97 & europe & ( 1.84, 0.38) & ( 0.75, 0.10) & ( 0.57,-0.01) & ( 0.36,-0.09) & ( 0.22,-0.11) & ( 0.04,-0.58) & (-0.23,-1.48) & (-0.33,-0.29) & (-0.45, 0.19) & (-0.52, 0.51) & (-0.50, 0.62) & (-0.65, 1.07) \\ 
  98 & gatekeeping & ( 1.85,-0.72) & ( 1.35,-0.23) & ( 0.94, 0.20) & ( 0.57, 0.46) & ( 0.26, 0.65) & (-0.02, 0.88) & (-0.20, 1.03) & (-0.35, 0.63) & (-0.42, 0.14) & (-0.50,-0.16) & (-0.81,-1.02) & (-0.81,-1.30) \\ 
  99 & youseeingthisshit & ( 2.36, 1.33) & ( 1.68, 0.67) & ( 1.58, 0.10) & ( 0.98,-0.47) & ( 0.56,-0.63) & ( 0.26,-1.23) & ( 0.02,-0.85) & (-0.17,-0.42) & (-0.38,-0.05) & (-0.56, 0.33) & (-0.76, 0.86) & (-0.80, 1.02) \\ 
\end{tabular}
}
\end{table}

\begin{table}[!ht]
    \hspace{-25mm}
    \scalebox{0.6}{
    \begin{tabular}{rlllllllllllll}
  100 & Art & ( 2.06,-0.93) & ( 2.03,-0.72) & ( 1.65, 0.37) & ( 0.83, 0.72) & ( 0.51, 0.87) & ( 0.12, 1.17) & (-0.10, 0.99) & (-0.29, 0.72) & (-0.41, 0.21) & (-0.55,-0.24) & (-0.68,-0.84) & (-0.79,-1.33) \\ 
  101 & confession & (-1.60, 0.63) & (-1.41, 0.41) & (-0.87,-0.07) & (-0.53,-0.39) & (-0.18,-1.00) & ( 0.20,-1.12) & ( 0.35,-0.77) & ( 0.48,-0.34) & ( 0.59, 0.11) & ( 0.78, 0.66) & ( 0.96, 1.14) & ( 1.04, 1.54) \\ 
  102 & awfuleverything & ( 3.39, 2.70) & ( 2.98, 2.11) & ( 2.29, 1.28) & ( 1.96, 0.75) & ( 1.16,-0.41) & ( 0.79,-0.97) & ( 0.48,-0.88) & ( 0.20,-0.70) & (-0.12,-0.37) & (-0.44, 0.07) & (-0.63, 0.53) & (-0.84, 0.99) \\ 
  103 & apple & ( 0.82, 1.75) & ( 0.79, 0.91) & ( 0.79, 0.76) & ( 0.77, 0.58) & ( 0.77, 0.35) & ( 0.71, 0.16) & ( 0.68,-0.04) & ( 0.61,-0.19) & (-1.09, 0.05) & ( 0.58,-1.05) & ( 0.67,-1.00) & ( 0.70,-0.92) \\ 
  104 & sports & ( 1.66,-0.66) & ( 1.08, 0.11) & ( 0.88, 3.07) & ( 0.22, 0.58) & ( 0.08, 0.76) & (-0.15, 0.79) & (-0.31, 0.79) & (-0.39, 0.39) & (-0.42, 0.01) & (-0.68,-0.54) & (-0.63,-0.68) & (-0.72,-0.86) \\ 
  105 & comedyheaven & ( 1.91, 1.53) & ( 1.74, 1.27) & ( 1.48, 0.90) & ( 1.17, 0.54) & ( 0.97, 0.06) & ( 0.53,-0.64) & ( 0.27,-0.85) & ( 0.03,-0.91) & (-0.22,-0.55) & (-0.50,-0.21) & (-0.81, 0.33) & (-1.12, 1.14) \\ 
  106 & entitledparents & ( 2.61, 1.46) & ( 2.59, 1.05) & ( 0.91,-2.03) & ( 0.32,-1.07) & ( 0.03,-0.68) & (-0.24,-0.41) & (-0.34,-0.16) & (-0.43, 0.12) & (-0.47, 0.34) & (-0.52, 0.52) & (-0.59, 0.76) & (-0.61, 0.89) \\ 
  107 & Overwatch & ( 1.63,-0.61) & ( 1.01,-0.08) & ( 0.69, 0.17) & ( 0.41, 0.38) & ( 0.22, 0.55) & ( 0.02, 0.92) & (-0.12, 0.97) & (-0.30, 0.67) & (-0.48, 0.17) & (-0.65,-0.51) & (-0.74,-1.06) & (-0.72,-0.93) \\ 
  108 & BikiniBottomTwitter & ( 1.36,-0.07) & ( 1.79,-0.80) & ( 0.74, 0.35) & ( 0.42, 0.49) & ( 0.18, 0.61) & (-0.11, 1.15) & (-0.27, 0.87) & (-0.43, 0.55) & (-0.56, 0.13) & (-0.63,-0.28) & (-0.71,-0.82) & (-0.85,-1.66) \\ 
  109 & atheism & ( 1.89,-0.72) & ( 1.14,-0.06) & ( 0.72, 0.21) & ( 0.59, 0.62) & ( 0.29, 0.56) & ( 0.01, 1.18) & (-0.15, 0.85) & (-0.31, 0.53) & (-0.41, 0.08) & (-0.53,-0.29) & (-0.67,-0.77) & (-0.73,-1.16) \\ 
  110 & anime & ( 1.61, 0.84) & ( 0.89, 0.30) & ( 0.69, 0.12) & ( 0.54,-0.02) & ( 0.30,-0.18) & ( 0.22,-0.90) & ( 0.00,-0.94) & (-0.25,-0.57) & (-0.43,-0.18) & (-0.56, 0.26) & (-0.72, 0.65) & (-0.91, 1.09) \\ 
  111 & TIHI & ( 3.32, 2.61) & ( 2.65, 1.55) & ( 1.93, 0.53) & ( 1.37,-0.28) & ( 1.09,-0.52) & ( 0.67,-1.03) & ( 0.42,-0.94) & ( 0.18,-0.76) & (-0.10,-0.40) & (-0.36, 0.01) & (-0.62, 0.49) & (-0.80, 0.93) \\ 
  112 & CasualConversation & ( 2.13, 0.93) & ( 1.51, 0.49) & ( 1.02, 0.18) & ( 0.75, 0.00) & ( 0.57,-0.14) & ( 0.27,-0.69) & ( 0.03,-1.38) & (-0.21,-0.51) & (-0.37,-0.06) & (-0.57, 0.31) & (-0.70, 0.65) & (-0.93, 1.21) \\ 
  113 & UpliftingNews & ( 1.98,-0.63) & ( 1.23, 0.05) & ( 1.01, 0.97) & ( 0.55, 1.13) & ( 0.17, 0.87) & (-0.08, 1.23) & (-0.23, 0.81) & (-0.36, 0.41) & (-0.49,-0.04) & (-0.52,-0.40) & (-0.63,-0.96) & (-0.68,-1.40) \\ 
  114 & dataisbeautiful & ( 0.01, 1.51) & ( 0.02, 1.10) & ( 0.10, 0.77) & ( 4.18,-0.42) & (-0.01, 0.54) & (-0.09, 0.56) & (-0.15, 0.49) & (-0.21, 0.28) & (-0.22, 0.10) & (-0.25,-0.18) & (-0.26,-0.46) & (-0.33,-1.63) \\ 
  115 & sex & ( 1.76,-0.84) & ( 1.32,-0.35) & ( 1.01, 0.05) & ( 0.71, 0.37) & ( 0.45, 0.60) & ( 0.14, 1.03) & (-0.05, 0.97) & (-0.25, 0.73) & (-0.45, 0.35) & (-0.59,-0.13) & (-0.75,-0.69) & (-0.92,-1.32) \\ 
  116 & EarthPorn & ( 2.04,-0.75) & ( 1.47,-0.25) & ( 1.09, 0.13) & ( 0.61, 0.47) & ( 0.40, 0.82) & (-0.02, 1.16) & (-0.17, 0.96) & (-0.32, 0.69) & (-0.49, 0.13) & (-0.63,-0.40) & (-0.69,-0.87) & (-0.88,-1.67) \\ 
  117 & whatisthisthing & ( 1.93,-0.49) & ( 1.12,-0.08) & ( 0.72, 0.14) & ( 0.42, 0.31) & ( 0.19, 0.52) & (-0.06, 1.21) & (-0.25, 1.16) & (-0.43, 0.54) & (-0.50,-0.06) & (-0.52,-0.49) & (-0.62,-0.91) & (-0.71,-1.25) \\ 
  118 & DunderMifflin & ( 1.60,-0.57) & ( 1.06,-0.13) & ( 0.81, 0.30) & ( 0.40, 0.48) & ( 0.19, 0.54) & (-0.08, 1.18) & (-0.29, 0.93) & (-0.43, 0.40) & (-0.56,-0.07) & (-0.69,-0.60) & (-0.74,-0.91) & (-0.82,-1.27) \\ 
  119 & LivestreamFail & ( 1.71, 0.90) & ( 1.11, 0.27) & ( 0.81, 0.04) & ( 0.60,-0.12) & ( 0.43,-0.24) & ( 0.22,-0.78) & ( 0.05,-0.96) & (-0.15,-0.69) & (-0.33,-0.22) & (-0.57, 0.22) & (-0.72, 0.76) & (-0.75, 0.92) \\ 
  120 & rareinsults & ( 4.15, 3.09) & ( 4.02, 2.68) & ( 2.64, 0.86) & ( 2.03, 0.00) & ( 1.18,-0.63) & ( 0.64,-1.03) & ( 0.39,-0.95) & ( 0.13,-0.71) & (-0.10,-0.35) & (-0.30, 0.04) & (-0.58, 0.53) & (-0.78, 1.00) \\ 
  121 & TwoXChromosomes & ( 1.85,-0.73) & ( 1.34, 0.02) & ( 0.93, 0.43) & ( 0.52, 0.88) & ( 0.21, 0.75) & (-0.04, 1.07) & (-0.24, 0.99) & (-0.34, 0.45) & (-0.48, 0.07) & (-0.56,-0.31) & (-0.63,-0.69) & (-0.76,-1.29) \\ 
  122 & 2meirl4meirl & ( 1.94, 0.75) & ( 1.50, 0.37) & ( 1.14, 0.02) & ( 0.74,-0.28) & ( 0.47,-0.42) & ( 0.15,-0.94) & (-0.07,-0.91) & (-0.34,-0.90) & (-0.44,-0.14) & (-0.57, 0.29) & (-0.71, 0.83) & (-0.87, 1.43) \\ 
  123 & instant\_regret & (-1.69, 0.44) & (-0.94, 0.03) & (-0.61,-0.17) & (-0.29,-0.23) & (-0.21,-0.49) & ( 0.15,-1.43) & ( 0.28,-0.78) & ( 0.42,-0.36) & ( 0.57, 0.03) & ( 0.68, 0.46) & ( 0.69, 0.79) & ( 0.88, 1.84) \\ 
  124 & niceguys & ( 1.88, 0.50) & ( 1.32, 0.13) & ( 0.87,-0.17) & ( 0.50,-0.42) & ( 0.26,-0.49) & (-0.06,-0.80) & (-0.32,-1.21) & (-0.42,-0.42) & (-0.49, 0.11) & (-0.63, 0.63) & (-0.66, 0.98) & (-0.74, 1.39) \\ 
  125 & Advice & ( 2.05,-0.95) & ( 1.80,-0.64) & ( 1.15,-0.02) & ( 0.81, 0.31) & ( 0.50, 0.57) & ( 0.19, 1.13) & ( 0.00, 1.07) & (-0.22, 0.74) & (-0.41, 0.31) & (-0.58,-0.18) & (-0.75,-0.75) & (-0.88,-1.26) \\ 
  126 & space & ( 1.83,-0.64) & ( 1.13,-0.06) & ( 0.85, 0.27) & ( 0.41, 0.25) & ( 0.18, 0.37) & ( 0.02, 1.06) & (-0.18, 0.93) & (-0.34, 0.65) & (-0.44, 0.19) & (-0.49,-0.21) & (-0.79,-1.09) & (-0.76,-1.35) \\ 
  127 & cats & ( 1.93,-0.94) & ( 1.36,-0.36) & ( 1.07, 0.05) & ( 0.71, 0.33) & ( 0.45, 0.53) & ( 0.19, 0.95) & ( 0.00, 0.94) & (-0.21, 0.70) & (-0.39, 0.30) & (-0.54,-0.15) & (-0.70,-0.66) & (-0.85,-1.23) \\ 
  128 & softwaregore & ( 1.82,-0.94) & ( 1.40,-0.40) & ( 1.21,-0.09) & ( 0.77, 0.28) & ( 0.49, 0.55) & ( 0.21, 0.93) & ( 0.01, 0.90) & (-0.22, 0.88) & (-0.38, 0.36) & (-0.62,-0.19) & (-0.76,-0.70) & (-0.92,-1.31) \\ 
  129 & hmmm & (-1.56,-0.54) & (-0.98, 0.07) & (-0.74, 0.55) & (-0.32, 0.68) & (-0.11, 0.83) & ( 0.16, 0.97) & ( 0.32, 1.01) & ( 0.42, 0.67) & ( 0.52, 0.30) & ( 0.61,-0.08) & ( 0.77,-0.73) & ( 0.94,-1.67) \\ 
  130 & fantasyfootball & ( 2.12, 2.10) & ( 2.32, 2.01) & ( 2.21, 1.47) & ( 2.12, 1.12) & ( 1.91, 0.74) & ( 1.55, 0.08) & ( 1.34,-0.78) & ( 0.61,-1.14) & (-0.27,-0.16) & (-0.37, 0.14) & (-0.36, 0.28) & (-0.42, 0.43) \\ 
  131 & ShitPostCrusaders & ( 1.90, 2.13) & ( 1.68, 1.68) & ( 1.44, 1.14) & ( 1.18, 0.55) & ( 0.91, 0.10) & ( 0.71,-0.53) & ( 0.48,-0.65) & ( 0.16,-0.62) & (-0.17,-0.37) & (-0.48,-0.04) & (-0.75, 0.34) & (-1.00, 0.75) \\ 
  132 & coolguides & ( 1.69, 2.64) & ( 1.18, 1.36) & ( 1.08, 1.15) & ( 0.95, 0.73) & ( 1.13, 0.67) & ( 0.88,-0.09) & ( 0.63,-0.42) & ( 0.45,-0.82) & ( 0.11,-0.61) & (-0.20,-0.49) & (-0.64,-0.08) & (-1.42, 0.82) \\ 
  133 & ATBGE & ( 1.95, 1.23) & ( 1.26, 0.51) & ( 1.08, 0.13) & ( 0.80,-0.36) & ( 0.46,-0.39) & ( 0.27,-1.02) & ( 0.04,-0.89) & (-0.17,-0.61) & (-0.34,-0.23) & (-0.61, 0.19) & (-0.78, 0.62) & (-1.08, 1.32) \\ 
  134 & Eyebleach & ( 2.06,-0.89) & ( 1.43,-0.24) & ( 1.12, 0.49) & ( 0.66, 0.72) & ( 0.38, 0.68) & ( 0.06, 1.13) & (-0.09, 0.98) & (-0.24, 0.70) & (-0.36, 0.32) & (-0.54,-0.09) & (-0.69,-0.61) & (-0.85,-1.39) \\ 
  135 & toptalent & ( 3.52,-1.45) & ( 2.23,-0.43) & ( 1.37,-0.05) & ( 1.16, 0.18) & ( 0.50, 0.51) & ( 0.22, 0.82) & ( 0.12, 0.74) & ( 0.00, 0.63) & (-0.12, 0.52) & (-0.22, 0.27) & (-0.41,-0.09) & (-0.85,-1.66) \\ 
  136 & worldpolitics & ( 1.83, 1.28) & ( 1.60, 0.84) & ( 1.92, 1.22) & ( 1.10, 0.12) & ( 0.78,-0.28) & ( 0.51,-0.69) & ( 0.35,-0.96) & ( 0.10,-0.87) & (-0.21,-0.75) & (-0.46,-0.24) & (-0.72, 0.32) & (-0.98, 1.08) \\ 
  137 & MovieDetails & ( 2.04, 1.12) & ( 1.57, 0.38) & ( 0.91, 0.05) & ( 0.63,-0.20) & ( 0.69,-0.65) & ( 0.29,-0.68) & ( 0.07,-0.91) & (-0.08,-0.70) & (-0.35,-0.44) & (-0.43, 0.04) & (-0.66, 0.57) & (-0.93, 1.36) \\ 
  138 & GetMotivated & ( 2.04,-0.64) & ( 1.16, 0.32) & ( 1.00, 2.26) & ( 0.42, 0.90) & ( 0.10, 1.71) & (-0.16, 1.39) & (-0.27, 0.77) & (-0.38, 0.30) & (-0.47,-0.07) & (-0.57,-0.57) & (-0.64,-0.85) & (-0.65,-1.05) \\ 
  139 & MapPorn & ( 1.50, 1.30) & ( 1.13, 0.75) & ( 1.07, 0.60) & ( 0.79, 0.29) & ( 0.74, 0.04) & ( 0.60,-0.60) & ( 0.25,-0.61) & ( 0.02,-1.14) & (-0.21,-0.39) & (-0.57,-0.07) & (-0.75, 0.36) & (-1.14, 0.97) \\ 
  140 & pcgaming & ( 1.72,-0.70) & ( 0.86, 0.24) & ( 0.57, 0.44) & ( 0.41, 0.86) & ( 0.35, 1.89) & (-0.11, 0.39) & (-0.23, 0.46) & (-0.38, 0.61) & (-0.40,-0.06) & (-0.57,-0.47) & (-0.58,-0.73) & (-0.60,-0.81) \\ 
  141 & wow & ( 1.42,-0.73) & ( 0.96,-0.24) & ( 0.68,-0.01) & ( 0.52, 0.19) & ( 0.44, 0.33) & ( 0.16, 0.80) & ( 0.04, 0.80) & (-0.22, 0.82) & (-0.53, 0.31) & (-0.66,-0.36) & (-0.75,-0.78) & (-0.77,-1.01) \\ 
  142 & reddeadredemption & (-1.33,-0.77) & (-0.93,-0.12) & (-0.63, 0.38) & (-0.41, 0.66) & (-0.17, 0.78) & (-0.05, 1.00) & ( 0.07, 0.95) & ( 0.19, 0.84) & ( 0.32, 0.72) & ( 0.58, 0.21) & ( 1.11,-1.07) & ( 0.96,-0.65) \\ 
  143 & im14andthisisdeep & ( 2.14, 1.40) & ( 1.62, 0.59) & ( 1.30, 0.02) & ( 0.83,-0.54) & ( 0.54,-0.82) & ( 0.28,-0.95) & ( 0.06,-0.84) & (-0.17,-0.61) & (-0.36,-0.26) & (-0.61, 0.24) & (-0.85, 0.81) & (-0.98, 1.22) \\ 
  144 & The\_Donald & (-1.44, 0.40) & (-0.64,-0.02) & (-0.45,-0.18) & (-0.26,-0.28) & (-0.11,-0.33) & ( 0.15,-1.37) & ( 0.36,-0.08) & ( 0.43, 0.13) & ( 0.48, 0.28) & ( 0.54, 0.42) & ( 0.57, 0.52) & ( 0.61, 0.62) \\ 
  145 & imsorryjon & (-1.44,11.07) & (-1.23, 9.62) & (-0.90, 5.07) & (-0.74, 2.96) & (-0.77, 1.03) & (-1.03,-0.17) & (-0.57,-0.27) & (-0.03,-0.12) & ( 0.41,-0.05) & ( 0.70, 0.02) & ( 1.10, 0.09) & ( 1.47, 0.15) \\ 
  146 & rarepuppers & ( 1.92,-0.67) & ( 1.23, 0.04) & ( 0.77, 0.64) & ( 0.45, 0.92) & ( 0.17, 0.75) & (-0.08, 1.08) & (-0.25, 1.05) & (-0.37, 0.60) & (-0.50, 0.08) & (-0.65,-0.45) & (-0.70,-0.93) & (-0.78,-1.43) \\ 
  147 & iamverybadass & (-1.47,-0.28) & (-0.91, 0.00) & (-0.46, 0.52) & (-0.14, 1.55) & ( 0.13, 0.56) & ( 0.37, 1.78) & ( 0.42, 0.70) & ( 0.55, 0.29) & ( 0.64,-0.19) & ( 0.66,-0.58) & ( 0.72,-0.91) & ( 0.70,-0.91) \\ 
  148 & fightporn & ( 3.08,-1.57) & ( 2.04,-0.52) & ( 1.51, 0.33) & ( 1.13, 1.35) & ( 0.72, 1.04) & ( 0.22, 1.21) & ( 0.04, 0.74) & (-0.16, 0.38) & (-0.33, 0.01) & (-0.45,-0.40) & (-0.49,-0.60) & (-0.59,-0.97) \\ 
  149 & gameofthrones & (-0.80, 1.46) & (-0.67, 1.14) & (-0.87, 1.26) & (-0.69, 0.79) & (-0.89,-1.09) & ( 0.68,-0.36) & ( 0.85,-0.04) & ( 1.04, 0.10) & ( 1.01, 0.14) & ( 1.03, 0.19) & ( 1.22, 0.23) & ( 1.13, 0.28) \\ 
  150 & youngpeopleyoutube & ( 3.77, 3.34) & ( 3.38, 2.76) & ( 2.43, 1.70) & ( 1.99, 0.99) & ( 1.11,-0.15) & ( 0.59,-0.79) & ( 0.35,-0.67) & ( 0.08,-0.47) & (-0.15,-0.19) & (-0.43, 0.16) & (-0.73, 0.58) & (-1.06, 1.13) \\ 
  151 & greentext & (-1.48, 0.48) & (-1.13, 0.11) & (-0.67,-0.17) & (-0.37,-0.38) & (-0.13,-0.45) & ( 0.16,-0.82) & ( 0.31,-0.89) & ( 0.43,-0.66) & ( 0.58,-0.13) & ( 0.68, 0.35) & ( 0.82, 0.86) & ( 0.97, 1.79) \\ 
  152 & TrueOffMyChest & ( 2.01,-1.11) & ( 1.62,-0.59) & ( 1.59,-0.34) & ( 1.03, 0.10) & ( 0.73, 0.22) & ( 0.51, 0.56) & ( 0.33, 0.71) & ( 0.08, 0.88) & (-0.23, 0.95) & (-0.44, 0.43) & (-0.68,-0.32) & (-0.93,-1.30) \\ 
  153 & offmychest & ( 2.16,-0.85) & ( 1.64,-0.19) & ( 1.19, 0.86) & ( 0.69, 1.38) & ( 0.26, 1.27) & (-0.12, 1.09) & (-0.23, 0.76) & (-0.39, 0.37) & (-0.46,-0.04) & (-0.56,-0.46) & (-0.63,-0.86) & (-0.70,-1.30) \\ 
  154 & BetterEveryLoop & ( 1.77,-0.77) & ( 1.24,-0.11) & ( 0.96, 0.64) & ( 0.56, 0.48) & ( 0.32, 0.54) & ( 0.07, 1.02) & (-0.10, 0.99) & (-0.29, 1.13) & (-0.45, 0.28) & (-0.81,-0.88) & (-0.77,-0.94) & (-0.73,-0.72) \\ 
  155 & tipofmytongue & ( 1.92, 1.22) & ( 1.29, 0.57) & ( 1.01, 0.29) & ( 0.70,-0.13) & ( 0.48,-0.38) & ( 0.34,-0.98) & ( 0.14,-1.07) & (-0.07,-0.78) & (-0.32,-0.40) & (-0.56, 0.08) & (-0.87, 0.64) & (-1.07, 1.22) \\ 
  156 & quityourbullshit & ( 1.74, 0.37) & ( 1.17, 0.09) & ( 0.72,-0.16) & ( 0.38,-0.25) & ( 0.16,-0.35) & (-0.15,-1.03) & (-0.32,-0.72) & (-0.56,-0.76) & (-0.57,-0.03) & (-0.60, 0.54) & (-0.64, 0.90) & (-0.76, 1.80) \\ 
  157 & hiphopheads & ( 1.53,-0.58) & ( 0.87, 0.22) & ( 0.47, 0.32) & ( 0.40, 0.55) & ( 0.17, 0.39) & (-0.06, 0.77) & (-0.24, 0.82) & (-0.30, 0.36) & (-0.39, 0.07) & (-0.51,-0.26) & (-0.52,-0.49) & (-0.67,-1.20) \\ 
  158 & woooosh & ( 1.76, 0.64) & ( 1.33, 0.20) & ( 0.95,-0.19) & ( 0.54,-0.57) & ( 0.25,-0.81) & (-0.09,-1.04) & (-0.29,-0.87) & (-0.45,-0.56) & (-0.58,-0.05) & (-0.69, 0.48) & (-0.82, 1.04) & (-0.92, 1.61) \\ 
  159 & ProgrammerHumor & ( 1.67,-0.94) & ( 1.09,-0.17) & ( 0.91, 0.18) & ( 0.53, 0.38) & ( 0.35, 0.56) & ( 0.18, 0.90) & ( 0.03, 0.83) & (-0.15, 0.73) & (-0.30, 0.40) & (-0.49, 0.06) & (-0.72,-0.52) & (-0.91,-1.18) \\ 
  160 & holdmyfeedingtube & ( 0.53, 1.02) & ( 0.52, 0.79) & ( 0.56, 0.88) & ( 4.25,-0.75) & ( 0.32, 0.83) & ( 0.05, 0.99) & (-0.07, 0.81) & (-0.21, 0.59) & (-0.25,-0.26) & (-0.31,-0.76) & (-0.32,-0.95) & (-0.35,-1.18) \\ 
  161 & nevertellmetheodds & ( 2.90, 0.98) & ( 2.24, 0.53) & ( 1.40,-0.21) & ( 1.00,-0.41) & ( 0.59,-0.82) & ( 0.13,-1.26) & (-0.08,-0.87) & (-0.27,-0.57) & (-0.40, 0.05) & (-0.55, 0.59) & (-0.51, 0.91) & (-0.57, 1.30) \\ 
  162 & CasualUK & ( 1.67,-0.75) & ( 0.97,-0.15) & ( 0.70, 0.01) & ( 0.86, 0.30) & ( 0.38, 0.27) & ( 0.10, 0.82) & (-0.08, 0.98) & (-0.30, 0.66) & (-0.38, 0.07) & (-0.57,-0.34) & (-0.60,-0.59) & (-0.78,-1.13) \\ 
  163 & CFB & ( 0.91,-1.22) & ( 0.98, 0.30) & ( 1.07, 0.56) & ( 1.00, 0.62) & ( 0.99, 0.72) & ( 0.76, 0.70) & ( 0.66, 0.70) & ( 0.23, 0.52) & (-0.22, 0.33) & (-0.35, 0.14) & (-0.58,-0.11) & (-0.69,-0.40) \\ 
  164 & shittysuperpowers & ( 0.02, 3.11) & ( 0.03, 2.42) & ( 0.00, 2.05) & ( 0.05, 2.12) & (-0.06, 0.69) & (-1.44,-0.30) & ( 0.07,-0.06) & ( 0.38,-0.14) & ( 0.56,-0.23) & ( 0.72,-0.32) & ( 0.84,-0.38) & ( 0.90,-0.43) \\ 
  165 & meme & ( 6.02,-2.26) & ( 4.68,-1.13) & ( 3.15,-0.04) & ( 2.00, 0.99) & ( 0.94, 1.76) & ( 0.30, 1.34) & ( 0.13, 1.05) & ( 0.00, 0.68) & (-0.15, 0.19) & (-0.25,-0.24) & (-0.35,-0.67) & (-0.43,-1.02) \\ 
  166 & thatHappened & ( 1.76,-0.72) & ( 1.14,-0.05) & ( 0.86, 0.58) & ( 0.57, 1.24) & ( 0.24, 0.99) & (-0.11, 1.09) & (-0.29, 0.78) & (-0.46, 0.34) & (-0.59,-0.12) & (-0.65,-0.56) & (-0.74,-0.98) & (-0.76,-1.26) \\ 
  167 & sadcringe & ( 2.23, 0.68) & ( 1.81, 0.28) & ( 1.17,-0.09) & ( 0.88,-0.31) & ( 0.51,-0.80) & ( 0.06,-1.35) & (-0.15,-0.87) & (-0.32,-0.47) & (-0.47, 0.06) & (-0.60, 0.66) & (-0.59, 0.88) & (-0.62, 1.34) \\ 
  168 & nonononoyes & ( 2.16,-0.75) & ( 1.35,-0.09) & ( 0.88, 0.20) & ( 0.51, 0.44) & ( 0.38, 0.84) & ( 0.05, 1.20) & (-0.12, 1.03) & (-0.30, 0.63) & (-0.41, 0.09) & (-0.59,-0.58) & (-0.63,-0.89) & (-0.77,-1.26) \\ 
  169 & technicallythetruth & ( 2.38,-1.50) & ( 1.69,-0.60) & ( 1.36,-0.13) & ( 1.02, 0.35) & ( 0.84, 0.62) & ( 0.50, 0.91) & ( 0.32, 0.89) & ( 0.04, 0.94) & (-0.20, 0.52) & (-0.46, 0.18) & (-0.70,-0.54) & (-0.92,-1.21) \\ 
  170 & MakeMeSuffer & ( 5.32, 4.57) & ( 4.92, 3.93) & ( 3.83, 2.44) & ( 2.52, 0.49) & ( 1.70,-0.28) & ( 1.16,-0.86) & ( 0.82,-1.05) & ( 0.45,-0.89) & ( 0.13,-0.60) & (-0.13,-0.23) & (-0.39, 0.20) & (-0.69, 0.80) \\ 
  171 & comics & (-1.89,-0.27) & (-0.57, 0.12) & (-0.31, 0.23) & (-0.14, 0.23) & (-0.04, 0.25) & ( 0.11, 0.41) & ( 0.13, 0.29) & ( 0.23, 0.25) & ( 0.25, 0.16) & ( 0.82,-1.62) & ( 0.40, 0.30) & ( 0.72, 1.34) \\ 
  172 & DotA2 & ( 1.34, 0.32) & ( 0.95, 0.32) & ( 0.75, 0.29) & ( 0.60, 0.19) & ( 0.44, 0.10) & ( 0.30,-0.46) & ( 0.03,-0.38) & (-0.37,-1.05) & (-0.42,-0.02) & (-0.49, 0.42) & (-0.59, 0.71) & (-0.65, 0.93) \\ 
  173 & confessions & (-1.75,-0.75) & (-0.99, 0.28) & (-0.81, 3.21) & (-0.11, 1.20) & ( 0.07, 0.89) & ( 0.32, 0.20) & ( 0.42, 0.07) & ( 0.50,-0.04) & ( 0.57,-0.18) & ( 0.63,-0.30) & ( 0.72,-0.46) & ( 0.74,-0.53) \\ 
  174 & TheMonkeysPaw & ( 2.18,-0.96) & ( 1.61,-0.43) & ( 1.15, 0.00) & ( 0.82, 0.30) & ( 0.48, 0.57) & ( 0.11, 1.16) & (-0.03, 0.95) & (-0.14, 0.55) & (-0.27, 0.25) & (-0.56,-0.18) & (-0.78,-1.03) & (-0.73,-0.93) \\ 
  175 & 2007scape & ( 1.46,-0.60) & ( 0.93,-0.14) & ( 0.69, 0.10) & ( 0.44, 0.31) & ( 0.23, 0.44) & (-0.03, 0.86) & (-0.17, 0.81) & (-0.33, 0.47) & (-0.44, 0.06) & (-0.55,-0.25) & (-0.64,-0.65) & (-0.77,-1.14) \\ 
  176 & conspiracy & ( 1.78, 0.64) & ( 1.25, 0.31) & ( 1.05, 0.15) & ( 0.72,-0.07) & ( 0.54,-0.12) & ( 0.25,-0.59) & (-0.01,-0.65) & (-0.31,-1.03) & (-0.36,-0.20) & (-0.52, 0.28) & (-0.65, 0.75) & (-0.75, 1.18) \\ 
  177 & baseball & ( 0.80,-0.16) & ( 0.71,-0.10) & ( 0.71,-0.15) & ( 0.64,-0.26) & ( 0.58,-0.21) & ( 0.42,-0.45) & ( 0.29,-0.30) & ( 0.17,-0.21) & (-0.02,-0.06) & (-1.29,-0.48) & (-0.28, 1.11) & (-0.09, 1.27) \\ 
  178 & depression & ( 2.17,-0.99) & ( 1.80,-0.51) & ( 1.29, 0.21) & ( 0.85, 0.78) & ( 0.46, 0.98) & ( 0.12, 1.16) & (-0.06, 1.05) & (-0.23, 0.71) & (-0.41, 0.24) & (-0.59,-0.30) & (-0.71,-0.78) & (-0.81,-1.24) \\ 
  179 & Instagramreality & ( 2.67, 1.03) & ( 1.96, 0.46) & ( 1.59,-0.05) & ( 1.00,-0.42) & ( 0.44,-0.63) & ( 0.06,-1.07) & (-0.11,-0.80) & (-0.26,-0.44) & (-0.40,-0.04) & (-0.47, 0.35) & (-0.55, 0.70) & (-0.70, 1.41) \\ 
  180 & borderlands3 & ( 2.46, 0.50) & ( 2.40, 0.45) & ( 2.36, 0.51) & ( 3.18, 0.55) & ( 1.69, 0.15) & ( 1.22, 0.07) & ( 0.97, 0.01) & ( 0.64,-0.24) & (-0.20,-0.90) & (-0.43, 0.15) & (-0.45, 0.98) & (-0.48, 1.39) \\ 
  181 & Cooking & ( 0.29, 0.87) & ( 0.18, 0.50) & ( 0.19, 0.36) & ( 2.73,-0.90) & ( 0.15, 0.24) & (-0.14, 0.12) & (-0.73,-1.50) & (-0.23, 0.19) & (-0.21, 0.33) & (-0.24, 0.48) & (-0.26, 0.56) & (-0.28, 0.61) \\ 
  182 & iamverysmart & (-1.44,-0.48) & (-1.04,-0.05) & (-0.79, 0.32) & (-0.35, 0.70) & (-0.15, 0.81) & ( 0.16, 1.32) & ( 0.37, 1.08) & ( 0.56, 0.79) & ( 0.64, 0.12) & ( 0.91,-0.98) & ( 0.83,-0.79) & ( 0.80,-1.19) \\ 
  183 & destiny2 & ( 1.67, 1.35) & ( 1.35, 0.75) & ( 1.17, 0.39) & ( 0.95, 0.01) & ( 0.74,-0.22) & ( 0.62,-0.79) & ( 0.42,-0.92) & ( 0.21,-0.82) & (-0.04,-0.57) & (-0.61,-0.21) & (-0.77, 0.51) & (-0.82, 0.84) \\ 
  184 & tumblr & ( 2.01,-0.80) & ( 0.83, 0.26) & ( 0.74, 0.82) & ( 0.35, 0.57) & ( 0.23, 0.63) & ( 0.02, 0.98) & (-0.11, 0.82) & (-0.22, 0.52) & (-0.35, 0.15) & (-0.49,-0.27) & (-0.59,-0.67) & (-0.70,-1.14) \\ 
  185 & gadgets & ( 1.81,-0.87) & ( 1.56,-0.59) & ( 0.87, 0.04) & ( 0.65, 0.27) & ( 0.57, 0.77) & ( 0.10, 0.92) & (-0.01, 0.85) & (-0.23, 0.80) & (-0.43, 0.38) & (-0.54, 0.02) & (-0.67,-0.43) & (-0.98,-1.52) \\ 
  186 & okbuddyretard & ( 1.96,-0.96) & ( 1.57,-0.41) & ( 1.19, 0.22) & ( 0.79, 0.61) & ( 0.50, 0.75) & ( 0.24, 0.98) & ( 0.06, 1.00) & (-0.14, 0.72) & (-0.33, 0.33) & (-0.47,-0.06) & (-0.61,-0.51) & (-0.75,-0.99) \\ 
  187 & AMA & ( 2.17,-1.32) & ( 1.56,-0.57) & ( 1.25,-0.04) & ( 1.03, 0.34) & ( 0.80, 0.54) & ( 0.54, 0.91) & ( 0.36, 0.96) & ( 0.02, 1.11) & (-0.26, 0.62) & (-0.47, 0.10) & (-0.75,-0.60) & (-0.86,-1.06) \\ 
  188 & ihadastroke & ( 2.51, 1.04) & ( 2.00, 0.66) & ( 1.66, 0.17) & ( 1.17,-0.24) & ( 0.65,-0.68) & ( 0.18,-1.08) & (-0.05,-0.99) & (-0.21,-0.57) & (-0.40,-0.09) & (-0.55, 0.38) & (-0.71, 0.98) & (-0.79, 1.38) \\ 
   \hline
\end{tabular}
}
\caption{Coordinates obtained using Correspondence Analysis for Subreddits with size at least $10^5$.}
\label{tab:CA_reddit}
\end{table}